\documentclass[twocolumn]{aastex63}
\usepackage{graphicx}
\usepackage{CJK}

\shorttitle{IR excess of OB stars}
\shortauthors{Deng et al.}

\graphicspath{{./}{Figures/}}

\begin{document}
\begin{CJK*}{UTF8}{gbsn}

\title{Infrared Excess of a Large OB Star Sample}

\correspondingauthor{Biwei Jiang}
\email{bjiang@bnu.edu.cn}

\author[0000-0003-0777-7392]{Dingshan Deng (邓丁山)}
\affiliation{Department of Astronomy, Beijing Normal University, Beijing 100875, China}
\affiliation{Lunar and Planetary Laboratory, University of Arizona, Tucson, AZ 85721, USA}

\author[0000-0001-6561-9443]{Yang Sun (孙漾)}
\affiliation{Department of Astronomy, Beijing Normal University, Beijing 100875, China}
\affiliation{Steward Observatory, University of Arizona, Tucson, AZ 85721, USA}

\author[0000-0001-5197-4858]{Tianding Wang (王天丁)}
\affiliation{Department of Astronomy, Beijing Normal University, Beijing 100875, China}
\affiliation{Dipartimento di Fisica e Astronomia Galileo Galilei, Universit`a di Padova, Vicolo dell’Osservatorio 3, I-35122 Padova, Italy}

\author[0000-0003-3860-5286]{Yuxi Wang (王钰溪)}
\affiliation{Department of Astronomy, Beijing Normal University, Beijing 100875, China}
\affiliation{College of Physics and Electronic Engineering, Qilu Normal University, Jinan 250200, China}

\author[0000-0003-3168-2617]{Biwei Jiang (姜碧沩)}
\affiliation{Department of Astronomy, Beijing Normal University, Beijing 100875, China}

\begin{abstract}
    
  The infrared excess from OB stars are commonly considered as contributions from ionized stellar wind or circumstellar dust.
  With the newly published LAMOST-OB catalog and GOSSS data, this work steps further on understanding the infrared excess of OB stars.
  Based on a forward modeling approach comparing the spectral slope of observational Spectral Energy Distributions (SED) and photospheric models, 1147 stars are found to have infrared excess from 7818 stars with good-quality photometric data. After removing the objects in the sightline of dark clouds, 532 ($\sim7\%$) B-type stars and 118 ($\sim23\%$) O-type stars are identified to be true OB stars with circumstellar infrared excess emission.
  The ionized stellar wind model and the circumstellar dust model are adopted to explain the infrared excess, 
  and Bayes Factors are computed to quantitatively compare the two.
  It is shown that the infrared excess can be accounted for by the stellar wind for about 65\% cases in which 33\% by free-free emission and 32\% by synchrotron radiation. 
  Other 30\% sources could have and 4\% should have a dust component or other mechanisms to explain the sharply increase flux at $\lambda > 10\mu$m. 
  The parameters of dust model indicate a large-scale circumstellar halo structure which implies the origin of the dust from the birthplace of the OB stars.
  A statistical study suggests that the proportion with infrared excess in OB stars increases with stellar effective temperature and luminosity, and that there is no systematic change of the mechanism for infrared emission with stellar parameters.

\end{abstract}



\section{Introduction} \label{sec:intro}


The infrared (IR) excess in early-type stars was firstly detected by \citet{Geisel_InfraredExcess_1970} who found  the observed color index $K$(2.2\,\micron)-$N$(10.2\,\micron) being redder than that corresponding to the star's spectral type. \citet{Geisel_InfraredExcess_1970} explained the infrared excess in the early-type stars by dust grains formed in the mass loss process like in the evolved low-mass stars. Then \citet{Allen_InfraredExcess_1972} detected several forbidden emission lines from some of these stars and argued that the $K-N$ excess comes from free-free emission by hot and ionized circumstellar gas instead of dust thermal emission.

\citet{hovhannessian_gasdust_2001Ap} studied 58 O, B, A, and F-type stars (including 45 OB stars) observed by the Infrared Astronomical Satellite (\textit{IRAS}). They explained the IR excess of 34 stars as the contribution by both blackbody emission from dust and free-free emission from ionized gas, and described the structure as `gas-dust shell' or `gas-dust disk'. Then, \citet{siebenmorgen_far-infrared_2018} identified twelve stars with IR excess from a sample of 22 OB stars with the \textit{Spitzer}/IRS spectrum available \citep{houck_infrared_2004} alongside the 2MASS \citep{skrutskie_2MASS_2006} and \textit{WISE} \citep{wright_wide-field_2010, cutri_vizier_2013} photometric data. Similar to the arguments of \citet{hovhannessian_gasdust_2001Ap}, the observational results can be successfully explained either by free-free emission from ionized gas or dust thermal emission.

The debate on the mechanism of infrared excess in early type stars has been continuing since its detection. Theoretically, early-type (OB) stars have strong stellar winds which produce hot and dense ionized gas to bring about IR excess by free-free emission \citep{hartmann_structure_1977}, which is a power-law continuum emission from IR to radio bands. Meanwhile, the IR excess can also be explained by optically thin dust emission. To discriminate the two mechanisms,  a wide-wavelength-range spectral energy distribution to well define the profile can be helpful. Nevertheless, previous works agreed that both free-free emission from ionized gas and thermal blackbody emission from dust can explain the infrared excess. But we are not clear how much proportion of the IR excess can be explained by free-free emission or dust emission and how the mechanism depends on stellar parameters. 

By its unique design of large field-of-view with four thousand of fibers, the Large Sky Area Multi-Object Fiber Spectroscopy Telescope (LAMOST; \citealt{cui_large_2012}) has acquired over 10 million stellar spectra in the Galaxy.
This huge database brings the possibility to significantly expand the scale of OB star sample.
From the LAMOST/LRS (Low Resolution Spectra), about 16,000 OB stars were identified by \citet{liu_catalog_2019} and from the LAMOST/MRS (Mid Resolution Spectra), other $\sim$9,000 OB stars were then identified by \citet{LAMOST_OBstars_MRS_Guo_2021}. 
For these LAMOST-OB stars, the basic stellar parameters including effective temperature ($T_{\mathrm{eff}}$) and surface gravity ($\log\,\mathrm{g}$) were determined by \citet{LAMOST_OBstarParameters_Guo_2021} by the data-driven technique Stellar LAbel Machine (SLAM) with the non-LTE TLUSTY synthetic spectra as the training dataset.
Moreover, the 1st catalog based on LRS is used by \citet{deng_intrinsic_2020} to determine the intrinsic color indexes of these OB stars and consequently an accurate measurement of the interstellar extinction to them, therefore the infrared excess can be calculated with high precision.

Based on this biggest sample of OB stars ever, this work tries to solve the following problems. How many of them have IR excess? What are their typical emission characteristics? How much do stellar wind and dust contribute?  The data will be introduced in Section~\ref{sec:Data} followed by the method to detect IR excess in the objects in Section~\ref{sec:IE}.
To explain this phenomenon, the models of ionized stellar wind and circumstellar dust will be presented in Section~\ref{sec:Model}.
The results are shown in Section~\ref{sec:Results} and more details will be discussed in Section~\ref{sec:Discussion}.

\section{Data}
\label{sec:Data}

\subsection{The OB star Sample}

The preliminary catalog contains more than 20,000 OB stars with stellar parameters by \citet{LAMOST_OBstarParameters_Guo_2021} from the LAMOST survey \citep{cui_large_2012} that is a reflective Schmidt telescope located at the Xinglong Station of the National Astronomical Observatory of China. 
In order to obtain more uniform properties from the statistical sample, and most of the parent samples are dwarfs, so only the dwarfs with $\log\,\mathrm{g} \ge 3.5$ is included in our sample.
Because this catalog lacks O-type stars, the Galactic O-Star Spectroscopic Survey (GOSSS, a project dedicated to O-stars; \citealt{maiz_apellaniz_galactic_2016}) which identifies more than 1000 O-stars is supplemented.

\subsection{Photometric Data}

The photometric data to define the spectral energy distribution (SED) covers the optical-to-infrared wavebands. The optical photometry from the \textit{Gaia} EDR3 \citep{gaia_collaboration_2016A&A...595A...1G,gaia_collaboration_2020yCat.1350....0G}, Pan-STARRS1 DR2 \citep{chambers_pan-starrs1_2016} and APASS DR9 \citep{henden_vizier_2016} surveys were adopted.
\textit{Gaia} takes high-quality photometry in three passbands: $G$, $G_{BP}$ and $G_{RP}$.
Pan-STARRS1 uses a 1.8\,m telescope located in Hawaii to observe in five bands: $g$, $r$, $i$, $z$ and $y$, and APASS works in traditional $B$ and $V$ bands.
In infrared, the data from 2MASS, \textit{Spitzer}/SEIP (\textit{Spitzer} Enhanced Imaging Products), \textit{Spitzer}/GLIMPSE \citep{churchwell_spitzerglimpse_2009} and \textit{WISE} surveys were adopted.
2MASS is an infrared full-sky survey in the $J$, $H$ and $K_s$ bands.
Both \textit{Spitzer}/SEIP and \textit{Spitzer}/GLIMPSE perform photometry in \textit{Spitzer}/IRAC (Infrared Array Camera; \citealt{fazio_spitzerIRAC_2004}) bands at 3.6, 4.5, 5.8, 8\,\micron\ respectively, and the 24\,\micron\ data from \textit{Spitzer}/SEIP and \textit{Spitzer}/MIPS (Multiband Imaging Photometer; \citealt{Rieke_SpitzerMIPS_2004}) are searched as well.
\textit{WISE}, though not as sensitive as \textit{Spitzer}/GLIMPSE, surveyed all the sky in the $W1$, $W2$, $W3$ and $W4$ bands.

In addition, the photometry at mid- and far-infrared wavelengths is searched from the observations by the \textit{MSX} satellite \citep{MSX_Price_2001AJ....121.2819P}, which surveyed the entire Galactic plane within $|b| \le 5\,\arcdeg$ in four mid-infrared spectral bands between 6 and 25\,\micron , by the \textit{IRAS} satellite \citep{Neugebauer1984_IRASpaperApJ...278L...1N} which surveyed more than 96\% of the sky at 12, 25, 60 and 100\,\micron\  as the first infrared space telescope, by the \textit{AKARI} satellite \citep{AKARI_Murakami2007PASJ...59S.369M} which covered more than 90\% of the sky at 9, 18, 65, 90, 140 and 160\,\micron, and by the \textit{Herschel Space Observatory}\citep{Pilbratt2010_HerschelSOA&A...518L...1P} which mapped nearly 8\% of the far-infrared up to sub-millimeter sky as the latest infrared space facility.
The OB stars are cross-identified in the \textit{IRAS} \citep{helou_IRAS_1988}, \textit{AKARI} \citep{AKARI_Ishihara2010A&A...514A...1I} and \textit{Herschel}/PACS point source catalogs \citep{HerschelPSC2020yCat.8106....0H}.
This results in 24 objects with IR excess which is detected in at least one band among these long wavelengths, which will be exclusively discussed later.
Consequently, the SED of most of the sample stars extends from optical to about 22-24\,\micron.

The cross-identification between catalogs are performed within a radius of 3$\arcsec$  (extending to 5$\arcsec$ for \textit{MSX}, \textit{IRAS}, \textit{AKARI} and \textit{Herschel}), which is about three times the positional uncertainties of LAMOST.
In cases where there is more than one object within the 3$\arcsec$ radius, the nearest one is selected.
The object is required to have the \textit{Gaia}, 2MASS and \textit{WISE} data available for a wide coverage of the SED. With these high-sensitivity all-sky surveys, the final LAMOST sample contains 20,551 stars, i.e. about 95\% of the preliminary sample, while the GOSSS sample is left with 589 sources, i.e. about 60\%.

\section{Detection of Infrared Excess}
\label{sec:IE}

\subsection{Extinction Correction}

Prior to searching for infrared excess, the interstellar extinction and reddening are corrected to get stellar intrinsic SED.
Since \textit{Gaia} provides the best photometry quality, the intrinsic color index \textit{Gaia}/$G_{BP}-G_{RP}$ is calculated by its relation with $T_{\mathrm{eff}}$ for early-type stars derived by \citet{deng_intrinsic_2020}, which brings about an accurate determination of the color excess $E(G_{BP}-G_{RP})$ that is proportional to absolute interstellar extinction.
The interstellar extinction in each photometric band is then calculated by the extinction law.
Specifically, this color excess is first converted to the classical reddening parameter $E(B-V)$ with $E(B-V)/E(G_{BP}-G_{RP}) = 0.757$ \citep{Wang_GalacticCCT_2018}, and then to extinction in other bands according to the extinction law by \citet{fitzpatrick_analysis_2007} in optical and by \citet{xue_precise_2016} in infrared.

The interstellar extinction is checked by comparing $E(B-V)$ with the widely used Bayestar 3D extinction map by \citet{green_3d_2019} at the \textit{Gaia} distance of the object.
For most of the objects, the two extinctions agree with each other very well. 
But they differ for about 10\% stars by $\bigtriangleup E(B-V)> 0.3$ mag, in particular for the stars with heavy reddening.
This is because \citet{green_3d_2019} take the average extinction in a range of area while our estimation dealt with the photospheric properties for individual star.
Thus, the interstellar extinction is corrected according to the intrinsic color index by \citet{deng_intrinsic_2020}.
One may question that the circumstellar dust can also redden the star and be taken into the interstellar extinction.
This may be true. 
But no correction is made to this point, because (1) we have no idea how much this might be, and (2) as will be shown later, this is a very small value on the order of 0.001 and negligible.

The photospheric emission should be calculated in order to extract the infrared excess.
The stellar atmospheric model from the Kurucz ATLAS9 \citep{Kurucz_ATLAS9_1979, castelli_notes_1997} and Tlusty \citep{hubeny_non-lte_1995, lanz_grid_2003, lanz_grid_2007} grids are  both examined.
Though the Tlusty model is more suitable for the massive stars such as OB stars, its grid begins from $T_{\mathrm{eff}} = 15,000$\,K.
Instead, the ATLAS9 model includes the range of $T_{\mathrm{eff}} \in [10000, 15000]\,$K. In addition, no apparent difference appears in these two models. So, the ATLAS9 model is adopted for all the sources.

The output flux from the stellar atmospheric model is converted to the observed flux by a normalization factor $C$:

\begin{equation}
  \label{Eq1:Fobs}
  C = (\frac{D}{R})^2 = \frac{F_{\mathrm{ATLAS9}}}{F_{\mathrm{obs}}}
\end{equation}

where $R$ refers to the radius of the stellar photosphere and $D$ is the distance.
For this purpose, only the optical bands of the \textit{Gaia}, APASS and PS1 missions are adopted because the infrared emission may come from circumstellar matter in addition to the photosphere.
In practice, the coefficient $C$ is calculated for each band and the mean value of $C$ is finally adopted.
The dispersion of $C$ from about 10 bands is typically about 10\%.
Together with the distance $D$ measured by \textit{Gaia} \citep{Bailer-Jones_gaia_distance_2021AJ....161..147B}, the stellar radius is derived by this mean $C$ and will be used later. 
Besides of $C$, the typical fractional uncertainty of $D$ is about 10\%. 
Therefore, based on the error propagation, the fractional uncertainty of stellar radius is also about 10\%.

Since $T_{\mathrm{eff}}$ and $\log\,\mathrm{g}$ are already given in the LAMOST-OB catalog, the closest model was simply matched to each star.
As for stars from GOSSS, the best model was selected by the least-$\chi^2$ method from all ATLAS9 grid for O-type stars ($T_{\mathrm{eff}} > 32000$\,K) and a pair of $T_{\mathrm{eff}}$ and $\log\,\mathrm{g}$ is given.

\begin{figure*}
  \gridline{\fig{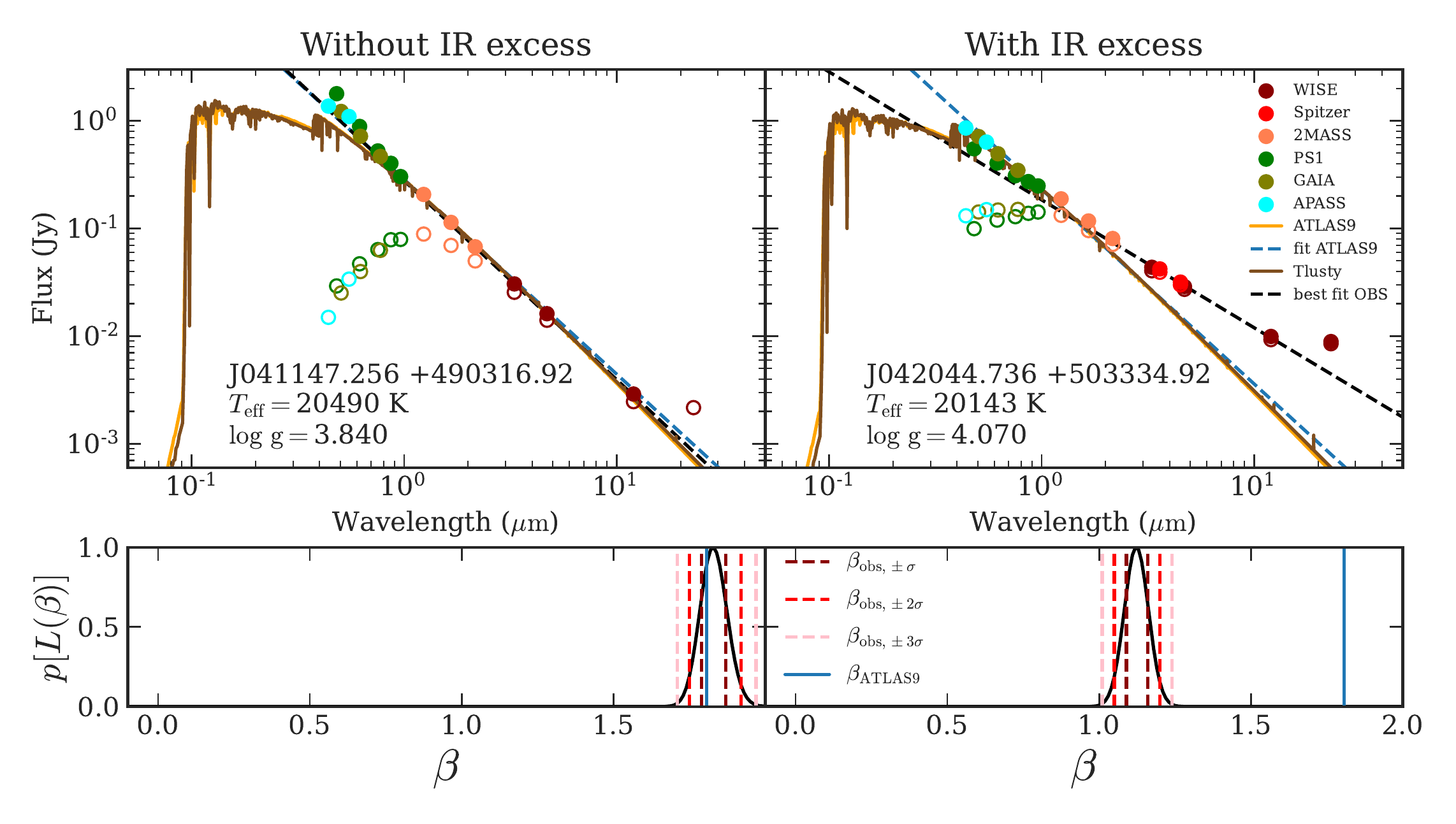}{0.87\textwidth}{}
            }
  \caption{Two typical spectral energy distributions for a star of $T_{\mathrm{eff}} \sim 20,000\,\mathrm{K}$ without (left) or with IR excess (right).
  The photometric data from a few surveys are presented, where the open and filled circles denote the flux before and after correction for interstellar extinction respectively.
  The photometry error is plotted in grey shades behind the data points.
  The orange and brown lines show the ATLAS9 and Tlusty models respectively for the given effective temperature and surface gravity from the LAMOST survey. 
  The dashed black and blue lines show the best-fit power-law $(F_{\mathrm{obs}} - F_{\mathrm{model}}) \propto \nu^{\beta}$ for observational SED and ATLAS9 model, respectively.
  The likelihood distribution regarding to spectral index $\beta$ is presented in the lower panels. 
  The dashed lines demonstrate 68\% (1$\sigma$), 95\% (2$\sigma$) and 99.7\% (3$\sigma$) confidence intervals and the solid blue lines represent the spectral slope index for the ATLAS9 photospheric data}
  \label{Fig1:2exampleSED}
\end{figure*}

Figure~\ref{Fig1:2exampleSED} shows two examples, including the photometric brightness before (open circle) and after (filled circle) correcting for interstellar extinction, and the model spectrum with the stellar parameters from the LAMOST spectroscopy.
Both the ATLAS9 and Tlusty models are displayed with no visible difference.

\subsection{Identifying Infrared Excess by Forward Modeling}
\label{subsec:IREidentify}

The aforementioned early detection of infrared excess in early type stars by \citet{Geisel_InfraredExcess_1970} used the color-index $K$(2.2\micron)-$N$(10.2\micron), while \citet{siebenmorgen_far-infrared_2018} detected the infrared excess by the ratio between observed flux and the blackbody photosphere flux as $\frac{F_{\mathrm{IRS}} - F_{\mathrm{BB}}}{F_{\mathrm{BB}}} > 0.1$ to find out the IR excess among their samples.
These two methods are simple to use, but strongly affected by the uncertainties of specific photometry data adopted for IR excess identification.

To avoid such issue, instead of doing a simple comparison on flux, a power-law fit on the long wavelengths SED is carried out with a spectral slope index $\beta$ (also see Figure~\ref{Fig1:2exampleSED}):

\begin{equation}
    F_\nu \propto \nu^{\beta}
\end{equation}

The IR excess could be detected by comparing the spectral slope index $\beta_\mathrm{obs}$ measured from the observational SED with the $\beta_\mathrm{ATLAS9}$ from ATLAS9 photospheric model.
Because the $T_\mathrm{eff}$ of our sources ranges from 10,000\,K to nearly 50,000\,K, the spectral slope of their theoretical photosphere radiation varies from approximately $\beta_\mathrm{ATLAS9} \sim 1.8$ to $\sim 2.0$, a universal threshold simply comparing these two $\beta$ is not appropriate. 
Hence, a forward modeling method based on Bayesian statistic framework is adopted.

The goodness of model fitting is calculated by the following likelihood function $L = -\exp{(\frac{\chi^2_{\mathrm{IR}}}{2})}$ which takes only the measurements at $\lambda > 2.15\,\micron$ (2MASS/$K_s$ band) into account, since the emission at shorter wavelength comes from stellar photosphere.
The $\chi^2_{\mathrm{IR}}$ is taken as:
\begin{equation}
  \label{Eq2:Chi2Equation}
  \begin{array}{ll}
       \chi^2_{\mathrm{IR}} &= \frac{1}{N} \sum_{i=1}^N \left(\frac{m_i^{\mathrm{model}} - m_i^{\mathrm{obs}}}{\mathrm{err}(m_i^{\mathrm{obs}})\times 2.5/\ln{10}}\right)^2 \\
       &= \frac{1}{N} \sum_{i=1}^N \left(\frac{\lg(F_i^{\mathrm{model}} / F_i^{\mathrm{obs}})}{\mathrm{err}(F_i^{\mathrm{obs}}) / F_i^{\mathrm{obs}}}\right)^2
  \end{array}
\end{equation}

where $m_i^{\mathrm{model}}$ and  $m_i^{\mathrm{obs}}$ are the model and observation photometry in magnitude, and  $F_i^{\mathrm{model}}$ and  $F_i^{\mathrm{obs}}$ are the model and observation flux, respectively. 
The synthetic photometry is also done from the modeling spectra to get the $F_i^{\mathrm{model}}$.

Quality control of the observational data is firstly conducted for accurate photometry data. The photometric error is limited to be smaller than 0.03\,mag, 0.05\,mag, 0.05\,mag and 0.1\,mag in the 2MASS/$K_s$, \textit{WISE}/$W1$, $W2$, and \textit{WISE}/$W3$ band respectively, which keeps 7671 sources in the LAMOST sample and 147 sources in the GOSSS sample.

Then, the infrared excess detection is carried out in following steps:

\begin{enumerate}
    \item For each star, the spectral index of its photospheric model $\beta_{\mathrm{ATLAS9}}$ is computed by least-square method from the $\lg F_\mathrm{ATLAS9} \propto \beta_\mathrm{ATLAS9} \times \lg \nu$. Synthetic photometry on the ATLAS9 model is performed based on the filters of each passband to get $F_{\mathrm{ATLAS9}}$, and no uncertainties are assumed.
    
    \item A grid of $\beta_{\mathrm{obs}}$ points ranging from $-5.0$ to $2.0$ with a step size of $0.01$ is created for finding the best one for observational data. For each of the $\beta_{\mathrm{obs, t}}$ value, an additional simulated flux $F_\mathrm{add} = C_\mathrm{add} \nu^{\beta_{\mathrm{obs, t}}}$ is added to the photospheric model to get the full modeling flux $F_\mathrm{model} = F_\mathrm{ATLAS9} + F_\mathrm{add}$. To best demonstrate any potential infrared excess at long wavelengths, the constant $C_\mathrm{add}$ is determined by forcing the $F_\mathrm{model}$ to match the \textit{WISE}/$W3$ observation.
    
    \item A distribution of likelihood function $L(\beta_\mathrm{obs})$ is compiled by going through the $\beta_{\mathrm{obs, t}}$ in the grid. The $\beta_{\mathrm{obs, t}}$ that gives the highest likelihood value is the best-fit point.
    
    \item Then, the 99.7\% confidence interval, corresponding to the 3$\sigma$ range in Gaussian distribution ($\beta_\mathrm{obs, -3\sigma} < \beta_\mathrm{obs} < \beta_\mathrm{obs, +3\sigma}$), is computed from the normalized likelihood distribution $L_\mathrm{norm} = L/(\int L(\beta_\mathrm{obs}) d\beta_\mathrm{obs})$.
    
    \item If the spectral slope from ATLAS9 photospheric model is outside of the 99.7\% confidence interval, i.e., $\beta_\mathrm{ATLAS9} > \beta_\mathrm{obs, +3\sigma}$, this star is recognised as having infrared excess.
    
\end{enumerate}

Top panels of Figure~\ref{Fig1:2exampleSED} show two stars as examples of the typical SEDs with or without IR excess.
The best-fit spectral index on the observations and ATLAS9 are shown as the dashed black and blue lines respectively. 
The $\beta_{\mathrm{obs}}$ and its likelihood distribution for these two examples are presented in the lower panels. 
The dashed lines demonstrate the boundaries of different confidence intervals and the solid blue lines show the spectral slope index for the ATLAS9 photospheric data. 
Smaller percentage for confidence intervals such as 68\% (1$\sigma$) or 95\% (2$\sigma$) could result in more detections of IR excess, but most of those additional samples are highly likely to be mis-identifications as the $\beta_{\mathrm{ATLAS9}}$ goes into the uncertainty range. 
Besides, the stellar parameters of these massive stars are with high uncertainty ($\mathrm{err}(T_{\mathrm{eff}})/T_{\mathrm{eff}} \sim 10\% $), resulting in some uncertainties of $\mathrm{err}(\beta_{\mathrm{ATLAS9}}) \sim 0.05$, but this uncertainty is not considered in this work. 
Therefore, as a safe choice, the upper bound of 99.7\% confidence interval is adopted aiming for clear IR excess identification.

For the entire sample of 7818 stars, 1147 stars are identified with IR excess, giving a percentage of $\sim 15\%$.

\subsection{Spectral Lines}
\label{subsec:SpectralLines}

LAMOST provides low-resolution (R1800) and medium-resolution (R7500) spectra, from which the spectral lines could be measured and analyzed. 
The $H_{\alpha}$ line index was measured by integrating a continuum-subtracted flux within $12\,$\AA\ for the 1071 of 1088 stars with clear mid-IR excess and available spectrum from LAMOST sample.
The continuum was subtracted by a second-order polynomial fit using $30\,$\AA\ of data on either side of the $H_{\alpha}$ line. 
We found 238 of them ($\sim 22\%$) have strong emission ($H_{\alpha}$ line index $> 10$). Three kinds of profiles appear as single-peak emission, self-absorption in the center and center emission with wing-absorption, which are expected from a circumstellar disk.
For stars that have multiple-epoch observations, though the time variation of line profile is obvious, no periodicity is visible, which may indicate the change is not due to periodic phenomena such as binary, rotation or pulsation. But the observations are not numerous enough, we will not investigate further based on the presently available data.

\subsection{Association with Dark Clouds}

Born in a dusty environment, many OB stars at their youth are still immersed in their birthplaces with significant amount of dust.
It is possible that background sources bring about  the IR excess.
Though a single-temperature modified blackbody of $f_{\nu} \propto \nu^2 B_\nu (T)$ radiation cannot fit the SED, the contribution by some surrounding clouds causes additional infrared radiation.
Dark clouds (DCs) are nearby members of the densest and coldest phase in the Galactic interstellar medium, and represent the sites where stars are currently being born.
Early-type massive stars are young and likely to be associated with those dark clouds with gas temperatures $ > 10 \,\mathrm{K}$, or the so-called IR dark clouds \citep{DC_Bergin&Tafalla2007ARA&A..45..339B}.
Both the spectrum and photometry of these stars in the dark clouds are very likely to be affected, and the observed infrared excess may then come from the cloud instead of the circumstellar matter.

In order to exclude the objects associated with dark clouds, the above sources that show infrared excess are cross-matched with the catalog of dark clouds from the Atlas and Catalog of Dark Clouds\footnote{http://darkclouds.u-gakugei.ac.jp/} (\citealt{DC_Dobashi2005PASJ...57S...1D}; \citealt{DC_Dobashi2011PASJ...63S...1D}) based on the optical Digitized Sky Survey (DSS) and infrared 2MASS images.
It is found that 463 stars from LAMOST and 34 stars from GOSSS are in the sightlines of dark clouds whose infrared excess is very likely to be caused by the radiation of dust in the cloud.
They are excluded in further analysis of the mechanism for infrared emission. 
After removing these stars from the preliminary sample, 625 stars from LAMOST and 25 GOSSS stars (650 stars in total) are left for further study.

\section{Modeling the IR Excess}
\label{sec:Model}

\subsection{Stellar Wind Model}

As massive stars, OB stars normally blow out strong hot ionized stellar wind.
\citet{seaquist_nature_1973} pointed out that for the isotropic spherical condition, the flux distribution of the free-free emission of electrons in an ionized gas which is generally thinner in the outer region following the power law  $F_\nu \propto \nu^{\alpha}$, where spectral index $\alpha \in [-0.1, 2]$. Theoretically, $\alpha= 2$ corresponds to the optically thick case that simulates the blackbody radiation, while $\alpha= -0.1$ refers to extremely optically thin case.
\citet{wright_radio_1975} derived that ionized stellar wind under spherical isotropic isothermal expansion, will have a flux of $F_\nu \propto \nu^{0.6}$.
This conclusion was then adopted to describe the radiation property in infrared to radio bands by many works such as \citet{crowther_physical_2007} and \citet{fogerty_silicate_2016}.
Furthermore, \citet{barlow_radio_1979} summarized the observational results of free-free emission from early-type stars as $F_\nu \propto \nu^{0.7}$,  which was adopted by \citet{siebenmorgen_far-infrared_2018} to explain the IR excess of OB stars.  Apparently,  the results of \citet{wright_radio_1975} and \citet{barlow_radio_1979} are highly consistent, with the index $\alpha \sim 0.6-0.7$, implying a rather optically thin free-free emission.

In our model, the power law $F_\nu = C_{\mathrm{F}} \nu^{\alpha}$ is adopted, where the spectral index $\alpha$ and scaling constant $C_{\mathrm{F}}$ are varied to fit the observational IR excess of each star.
The upper limit of $\alpha$ is set to 2.0, coincide with both the Rayleigh-Jeans approximation of the high-temperature photospheric radiation and the free-free emission of the electrons. On the other hand, the lower limit of $\alpha$ is free. Though the lower limit of $\alpha$ is $-0.1$ for free-free emission as mentioned above, synchrotron radiation, which also follows a power law $F_\nu \propto \nu^{\alpha}$ but with a much larger negative index, i.e. $\alpha < -0.1$, is a possible source of infrared excess. \citet{Shchekinov&Sobolev2004_SynchrotronMassivestars_A&A...418.1045S} argue that the interaction of stellar wind with the surface of a circumstellar accretion (or protoplanetary) disk around massive stars can result in the acceleration of relativistic electrons in an external layer of the disk and produce synchrotron radiation. Leaving the lower limit of $\alpha$ free opens the possibility of identifying synchrotron radiation as the source of infrared excess. While the modelling simply takes the power law, the free-free and synchrotron radiation will be discriminated by the power law index yielded from the modelling.

Similar to the approach in Section~\ref{subsec:IREidentify}, a grid of $\alpha$ and $C_{\mathrm{F}}$ is created to compile the likelihood distribution of $L(\alpha, C_{\mathrm{F}})$.
The spectral index $\alpha$ is ranging from -10.0 to 2.0 with a step of 0.1, and 10 $C_{\mathrm{F}}$ grid points with uniform interval sampled between the value matching the lowest-flux photometry and the highest-flux at the wavelengths $> 2.15\,\micron$.
Then, the best-fit parameters are identified with the highest likelihood value and the uncertainties for both are determined from 68\% confidence intervals.

\subsection{Circumstellar Dust Model}
\label{subsec:dustymodel}

Other than the stellar wind, thermal radiation from circumstellar dust could also be responsible for the IR excess around massive stars.

The code DUSTY \citep{Ivezic_DUST_1997, Ivezic_DUSTY_1999} is adopted to analyze the properties of circumstellar dust.
DUSTY solves the radiative transfer problem in a dusty environment and offers many options for input radiation, dust types and density distributions.
Because for most of the stars in our sample, they are lack of long wavelengths observations, thus hard to constrain their dust parameters.
A simple Bayesian approach considering a prior distribution is adopted.

\begin{figure*}
\gridline{\fig{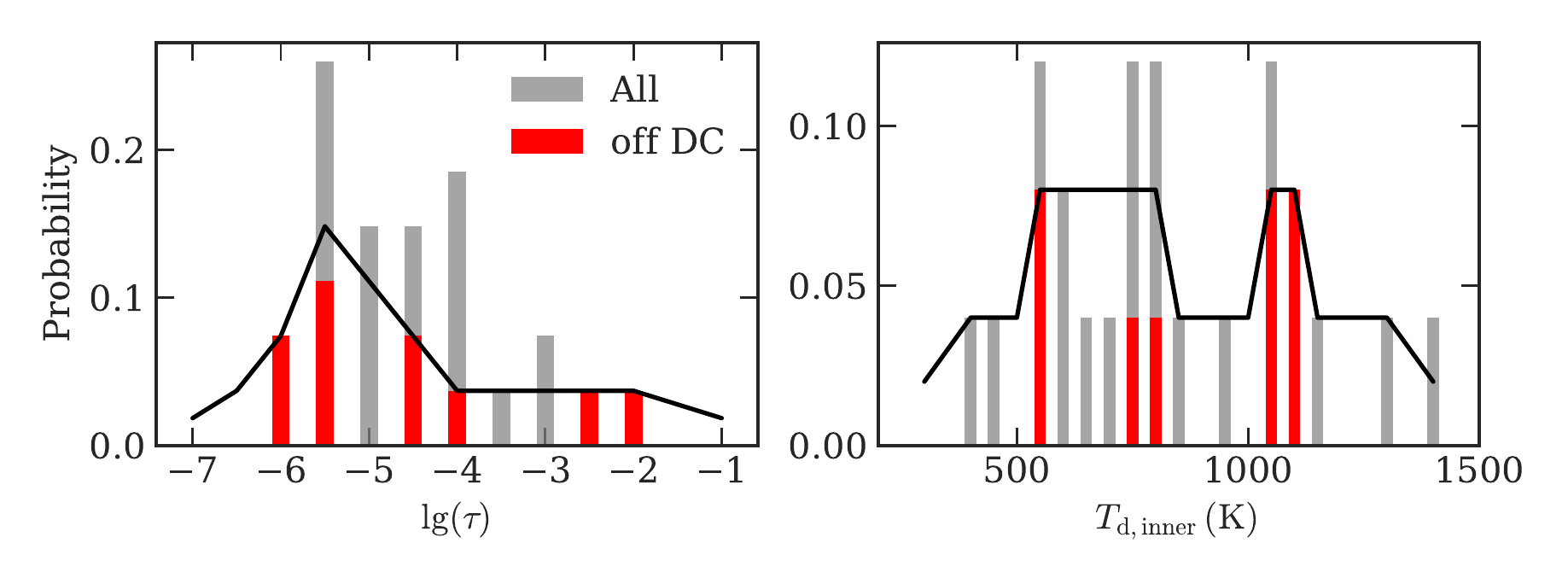}{0.87\textwidth}{}
          }
\caption{
  The prior probability distribution of dust parameters $\tau$ and $T_\mathrm{d}$ based on the 27 training samples with long wavelengths observations, so that the dust parameters can be better constrained. The columns are the parameters fitted best for individual star, while the red columns refer to the one off the sightlines of dark clouds. The black lines represent the selected prior probability distribution later used for analyzing the whole sample.
  }
\label{Fig2:Prior}
\end{figure*}

There are 27 stars that contain long wavelengths observations at $\lambda \ge 60$\,\micron\ from \textit{IRAS}, \textit{AKARI} or \textit{Herschel}, standing out from most of our samples that the observation only reaches to the wavelengths of $W3$ band of 12\,\micron. These stars are chosen as the training sample for dust model fitting to obtain the priors of dust parameters.
Using the training sample, the parameters of the DUSTY models library are chosen to cover the reasonable range for OB stars as following.
First, the isotropic central radiation of a single heat source is adopted, specifically the input stellar spectrum is taken from the ATLAS9 model at given stellar parameters.
Some dust-related parameters are fixed as well:  the size distribution follows the MRN power law i.e. $n(a) = a^{-q}$ for $a \in [0.005, 0.25]\,\micron$ and $q=3.5$; the upper limit of the dust temperature is set to be 1500\,K; the outer radius of this dust shell is set to be $10^3$ times of the inner radius.
The grids of models are built within a range of parameters.
For the temperature of inner dust shell ($T_{\mathrm{d, inner}}$, hereafter $T_{\mathrm{d}}$), 29 equally spaced points with an interval of 50\,K are sampled from 100\,K to 1,500\,K.
The optical depth at $550\,\mathrm{nm}$ ($\tau^{\rm dust}_{\rm V}$, hereafter $\tau$) is explored at 33 equal logarithmic interval from $10^{-7}$ to $10$.
Various combinations of chemical composition (e.g. silicate, amorphous carbon and graphite) and dust density distributions (e.g. inverse square attenuation with radius and the AGB stellar wind density model) are also tested.
The experimental running of DUSTY found that the optical depth of dust $\tau$ is very low that the discrimination of dust species is meaningless.
Thus, the dust composition is fixed as a mixture of 53\% silicate and 47\% graphite, i.e. the average interstellar dust composition from \cite{draine_optical_1984}.
This option presumes that the dust around these OB stars comes from their birthplace -- molecular clouds, which is evidenced by the sub-parsec-scale dust structure ($\sim 0.1\,$pc)  since stellar wind can hardly reach such a distance.
Correspondingly, the dust density distribution is set to be constant.
All physical parameters adopted are shown in Table~\ref{tab1:DustyPara}. 
Also from this training sample, a simple prior probability distribution of $\tau$ and $T_{\mathrm{d}}$ is set up, as shown in Figure~\ref{Fig2:Prior}.

\begin{deluxetable*}{cc}
\tablecaption{The parameters value and range adopted in the DUSTY model \label{tab1:DustyPara}}
\tablewidth{0pt}
\tablehead{
\colhead{parameter}  & \colhead{Value / Range}
}
\startdata
Geometry                    & Isotropic sphere \\
Input stellar spectrum      & ATLAS9  \\
Dust temperature in the inner radius ($T_{\mathrm{d, inner}}$)
                            & from 100\,K to 1500\,K with an equal interval of 50\,K \\
Chemical composition        & 47\% Silicate \& 53\% Graphite \citep{draine_optical_1984} \\
Density distribution        & constant with the outer radius $r_{\mathrm{out}} = 10^3 r_{\mathrm{inner}}$ \\
optical depth at $550\,\mathrm{nm}$ ($\tau^{\rm dust}_{\rm V}$) & 33 values from $10^{-7}$ to $10$ with equal logarithmic interval \\
\enddata
\end{deluxetable*}

From this training sample, a wind-dust model is established by combining the DUSTY model and the stellar wind model of $F_\nu = C_{\mathrm{F}} \nu^{\alpha}$ in following steps:

\begin{enumerate}
    \item Compute a grid of dust models from DUSTY. For each combination of $T_{\mathrm{d}}$ and $\tau$ from dust parameter grid described in Table~\ref{tab1:DustyPara}, dust thermal radiation is first calculated for each star as $F_{\mathrm{dust}}$.
    \item For each source, compute the residual for each combination of dust parameters ($T_{\mathrm{d}}, \tau$) and its photosphere, then fit the residual with a power-law by a simple least-squared method: $F_{\mathrm{wind}} = C_{\mathrm{F}} \nu^{\alpha} = F_{\mathrm{obs}} - F_{\mathrm{ATLAS}} - F_{\mathrm{dust}}$.
    \item Compile a distribution of likelihoods $L_\mathrm{test}(T_\mathrm{d}, \tau)$ by going through all the $T_{\mathrm{d}}$ and $\tau$ in the grid.
    \item Multiply the likelihoods with the prior found by training samples to get the posterior: $L_\mathrm{post}(T_\mathrm{d}, \tau) = L_\mathrm{test}(T_\mathrm{d}, \tau) P_\mathrm{prior}$. From this posterior distribution, the values $(T_\mathrm{d}, \tau)$ given highest probability is chosen as the best-fit parameters, and the uncertainties are also presented accordingly as 68\% confidence intervals.
\end{enumerate}

As the same as the above method, only photometry data with wavelength $> 2.15\,\micron$ are considered in computing $L_\mathrm{test}(T_\mathrm{d}, \tau)$ to exaggerate the differences among the grids.
Differently, the spectral index in this wind-dust model is limited within $\alpha \in [-0.1, 2]$ to represent free-free emission from the ionized stellar wind, while synchrotron radiation is not taken into account.
Though the stellar wind component is added to all stars, it is unnecessary for some of them for which $C_{\mathrm{F}}$ is very small.
The post likelihood distribution and the chosen fit parameters are highly sensitive to the prior distribution based on the training sample of 27 stars. But due to the lack of long wavelengths observations on those OB stars in the total sample, this is currently the best estimation that could be chosen on wind-dust model.

\subsection{Model Comparison}
\label{subsec:modelcomp}

To quantitatively understand which model explains observations better, Bayes Factors ($\mathrm{BF}$, \citealt{Jeffreys61_BFs, KassRafferty1995_BayesFactors}) are computed between two models for each star:

\begin{equation}
    \label{Eq3:BF}
    \mathrm{BF} = \frac{P({\cal M}_1\mid \mathrm{data})}{P({\cal M}_2\mid \mathrm{data})}
\end{equation}

where ${\cal M}_i$ refers to two different models and $P$ is the marginalized probability.
There are two grids as described above for each star: (a) stellar wind model with likelihood distribution $L(\alpha, C_{\mathrm{F}})$, (b) wind-dust model with $L(T_\mathrm{d}, \tau)$.
From these grids, the marginalized posterior probability  $P({\cal M}_i\mid \mathrm{data})$ is computed by:

\begin{equation}
    \label{Eq3:Pgl}
    P({\cal M}_i\mid \mathrm{data})  = \int \int L_\mathrm{post}(\lambda_1, \lambda_2) d\lambda_1 d\lambda_2
\end{equation}
where $\lambda_1$ and $\lambda_2$ refer to the two free parameters in two models, and $L_\mathrm{post}$ is the posterior probability.
The computed $\mathrm{BF}$ value can tell which model is better quantitatively between the two models.


\section{Results}
\label{sec:Results}

Previous works usually found that nearly 50\% of the sample OB stars show IR excess.
Excluding the stars in the dark clouds sightlines, there are 532/5634 ($\sim 9.4\%$) B-type and 118/301 ($\sim 39.2\%$) O-type stars with true infrared excess.
This apparently lower percentage of OB stars with IR excess may be attributed to the fact that our OB star sample is not biased to any objects potentially having IR excess, instead it mainly comes from the LAMOST survey in optical. Moreover, the infrared excess that may come from surrounding medium is exclusively removed. 
If taking the ratio of observed flux in the infrared band to the photosphere flux as \citet{siebenmorgen_far-infrared_2018} did, a larger proportion of IR excess would be obtained.
However, the additional stars usually have marginal IR excess. Our results may represent the portion with IR excess among normal OB stars.

\begin{figure}
  \gridline{\fig{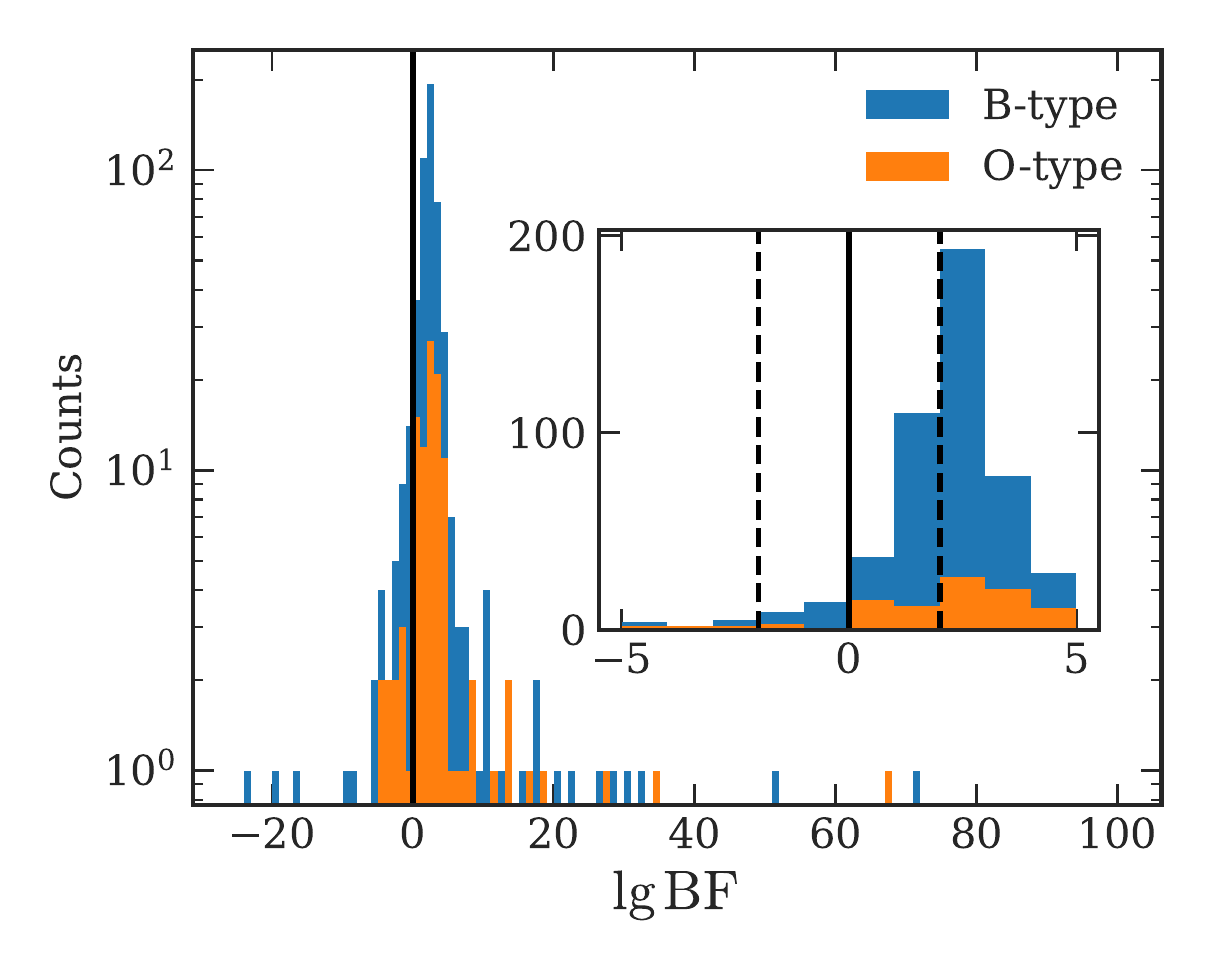}{0.47\textwidth}{}
            }
  \caption{
    The distribution of Bayes Factors ($\mathrm{BF}$), which is defined as overall goodness of stellar wind model over wind-dust model. B-type and O-type stars are represented in blue and orange histograms, respectively. The outer panel is shown in log scale while the inner one zooming in to the region of $\lg \mathrm{BF} = $ -5 to 5 is in linear scale. The $\lg \mathrm{BF} = 0$ line is shown in black line, and $-2$ and $2$ in dashed lines.
    }
  \label{Fig3:BFdist}
\end{figure}

The Bayes Factor ($\mathrm{BF}$, described in Section~\ref{subsec:modelcomp}) is computed for each star. 
In this work, the stellar wind model is called ${\cal M}_1$ while the wind-dust model called ${\cal M}_2$.
When the $\mathrm{BF} > 100$ (or $< 0.01$), which means ${\cal M}_1$ model is 100 times more (or less) likely to explain the observations than the ${\cal M}_2$ \citep{KassRafferty1995_BayesFactors}, the ${\cal M}_1$ (${\cal M}_2$) would be assigned as the model that best explains the data, and when $0.01 \le \mathrm{BF} \le 100$, both are appropriate models.
The $\mathrm{BF}$ distribution for all the stars are presented in Figure~\ref{Fig3:BFdist}.

Since the stellar wind component is also in the wind-dust model, ideally the likelihood distribution should be complied from $L(T_\mathrm{d}, \tau, \alpha, C_{\mathrm{F}})$ with four variables together. 
However, it is not practical as the computation time for each star with only two variables $(T_\mathrm{d}, \tau)$ is already $\sim$ 1 minute. 
If a typical grid with $\sim$ 100 points for $(\alpha, C_{\mathrm{F}})$ were added to the whole grids, the whole process would be too computational expensive. 
A Markov chain Monte Carlo (MCMC) simulation would be helpful, but it is not necessary here. 
Because the best stellar wind component ($\alpha_\mathrm{best}, C_{\mathrm{F, best}}$) is already included in the wind-dust model for each combination of the dust parameters ($T_\mathrm{d}, \tau$), the marginalized probability computed for wind-dust model here is larger than the marginalized probability from the comprehensive one: 
\begin{equation}
\begin{array}{ll}
     & \int L_\mathrm{best (\alpha, C_{\mathrm{F}})}(T_\mathrm{d}, \tau) dT_\mathrm{d} d\tau  \\
     &\sim \int L(T_\mathrm{d}, \tau, \alpha_\mathrm{best}, C_{\mathrm{F, best}}) dT_\mathrm{d} d\tau d\alpha dC_{\mathrm{F}} \\
     &\ge \int L(T_\mathrm{d}, \tau, \alpha, C_{\mathrm{F}}) dT_\mathrm{d} d\tau d\alpha dC_{\mathrm{F}}.
\end{array}
\end{equation}
With an overestimated marginalized probability for wind-dust model, the comparison would be inclined to it.
Even though, as it shows below, the overall comparison tells that only dozens of stars are clearly better explained by wind-dust model, while stellar wind model is still a better choice or at least as good as the wind-dust model for most of the stars in our sample. 
Thus, the adopted simpler grid with only two variables, $L(T_\mathrm{d}, \tau)$, would not give a different result comparing with the comprehensive one.

\begin{figure*}[tb]
  \gridline{\fig{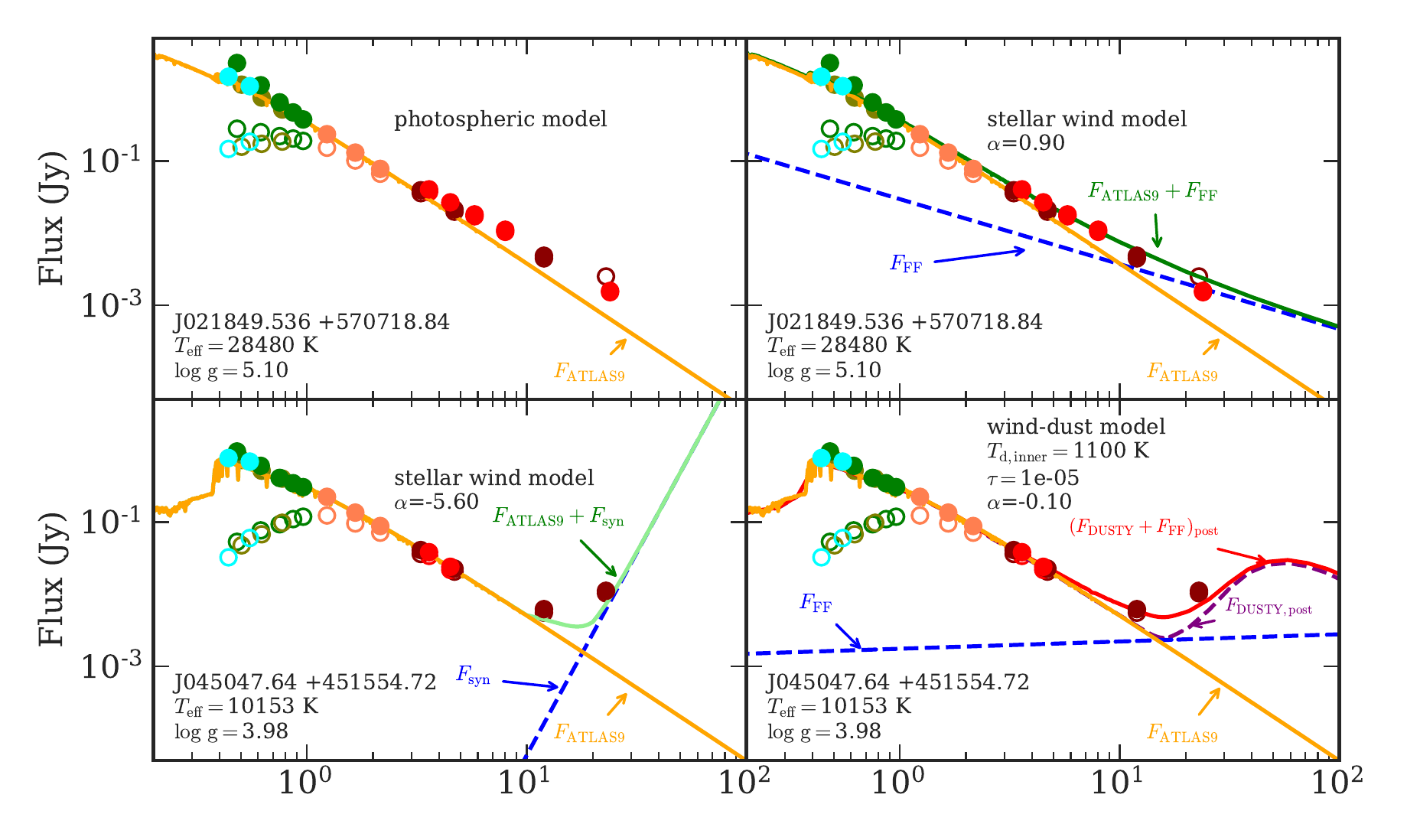}{0.87\textwidth}{}
            }
  \caption{
    Two typical cases of model fitting from top to bottom: (1) The SED can be better fitted by a stellar wind model (right) than a photospheric model; (2) The SED can be fitted by either a stellar wind model (left) or a wind-dust model (right). The symbols follow the convention in Figure~\ref{Fig1:2exampleSED}, while meanings of lines can be found in each panel.
    For the power-law radiation component represented in blue dashed lines, $F_\mathrm{syn}$ is for synchrotron radiation when $\alpha < -0.1$ and $F_\mathrm{FF}$ is for free-free emission when $\alpha \in [-0.1, 2]$.
    }
  \label{Fig4:SEDmodelExample}
\end{figure*}

\begin{figure}
  \gridline{\fig{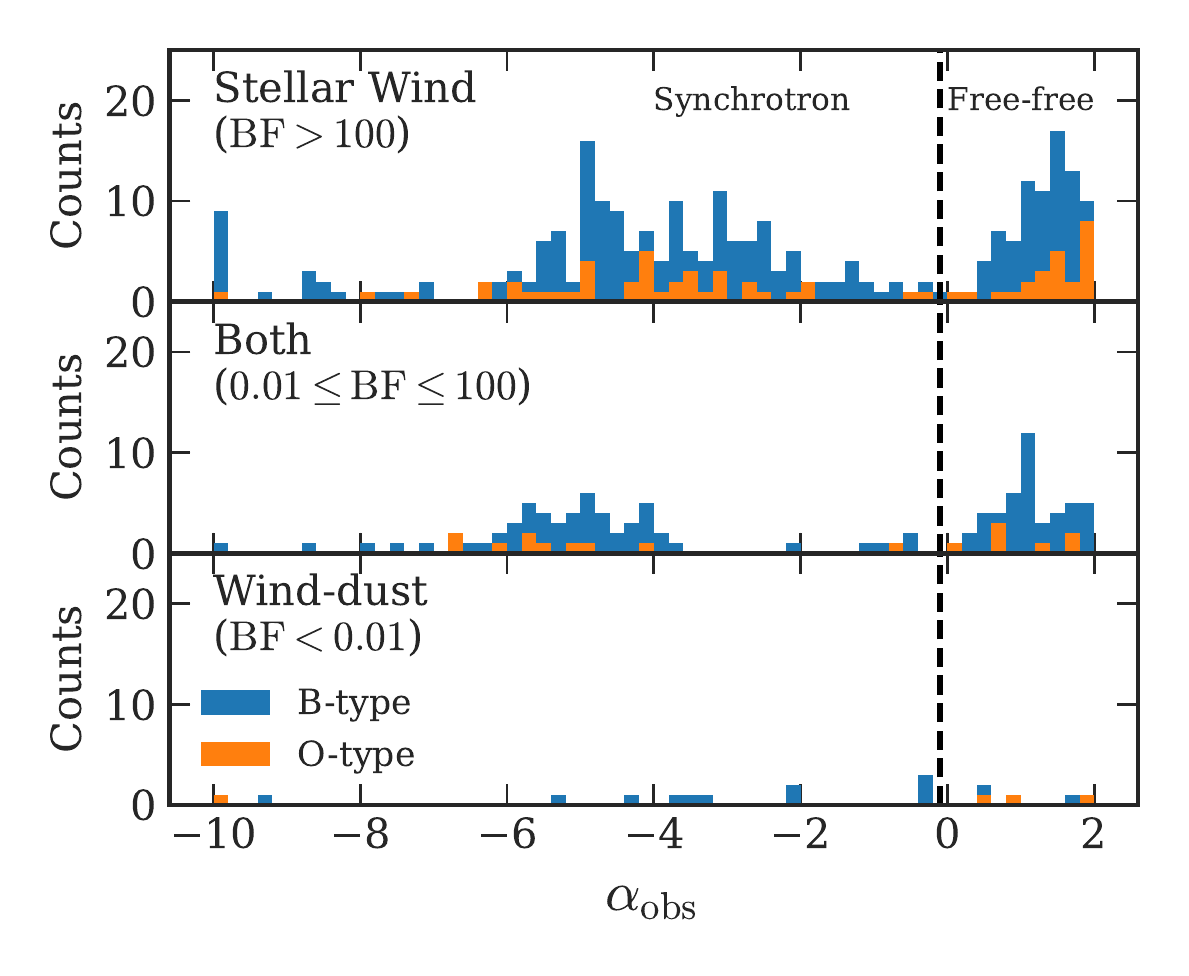}{0.47\textwidth}{}
            }
  \caption{
    The distribution of spectral index $\alpha$ in the stellar wind model for B-type (blue) and O-type (orange) stars.
    The samples that could be fitted best by stellar wind model, both models or wind-dust model are presented from top to bottom with the criterion based on $\mathrm{BF}$ noted in the text.
    The black line marks where $\alpha = -0.1$. There are two clear groups on either side of this line.
    }
  \label{Fig5:alphaDis}
\end{figure}

For 532 B-type and 118 O-type stars off the dark clouds sightlines and with IR excess, the stellar wind model with theoretical predicted free-free emission ($\alpha \in [-0.1, 2]$) can satisfactorily explain the infrared excess of 178 ($\sim 19\%$) and 39 ($\sim 19\%$) stars, respectively.
In addition, there are 169 ($\sim 18\%$) B-type and 40 ($\sim 19\%$) O-type stars, respectively for which steep increase at $\lambda> 10\,\micron$ can be fitted by a single power-law radiation, but requiring a larger negative spectral index ($\alpha < -0.1$), e.g. the star in the second row in Figure~\ref{Fig4:SEDmodelExample}.
Such large negative index is generally an indicator of synchrotron radiation. 
Figure~\ref{Fig5:alphaDis} shows the distribution of the spectral index $\alpha$ fitted in the stellar wind model for all the stars with IR excess and off dark clouds. 
There are clearly two groups on either side of $\alpha = -0.1$, suggesting there are two separate mechanisms.
In addition, there are 10 stars fitted by $\alpha = -10$ which is on the lower edge of the grid and the upper bound of 68\% confidence interval could even reach to $\alpha_{+\sigma} \sim 1.3$, giving a huge uncertainty of $\Delta \alpha > 10$.
It is because there is only one photometry data point (\textit{WISE}/$W3$) showing obvious IR excess in their SED. 
Hence, the constraint on the spectral index $\alpha$ is so weak that it is not even possible to pin down the best value.

It should be mentioned that when a star can be explained well by the power-law model no matter the value of the spectral index $\alpha$, it is possible that the IR excess comes from an exceptionally optically thin dust shell.
There are a clear group suggesting such trend in the second row of Figure~\ref{Fig5:alphaDis}, where both models fit the observational SED as good as each other.
Moreover, a dust component of $\tau < 10^{-7}$, which is even smaller than the lower limit in our grid could even fit the stars that are currently fitted better by stellar wind model.
A practical problem is that the observations available to constrain the dust model are scarce and lacking at long wavelengths, and an obligatory dust model would be unreliable.
In addition, for those with $\alpha \in [-0.1, 2]$, the free-free emission from stellar wind model already works quite well, and it is theoretically plausible.
Hence, the free-free emission from ionized stellar wind is set to be the best model for those stars with $\alpha \in [-0.1, 2]$ and $\mathrm{BF > 100}$.
From them, the mean spectral index in $F_\nu \propto \nu^{\alpha}$ can be concluded as $\left \langle \alpha \right \rangle = 1.603$ with a standard deviation of $\sigma_{\alpha} = 0.492$ for B-type stars, and $\left \langle \alpha \right \rangle = 1.857$ with a standard deviation of $\sigma_{\alpha} = 0.313$ for O-type.
This mean index for B-type stars of 1.603 is higher to the expected value of $\sim 0.7$ from \citet{wright_radio_1975} and \citet{barlow_radio_1979}, because the IR excess identification methodology adopted here is so sensitive that even stars with very weak IR excess are identified.

\begin{figure}
  \gridline{\fig{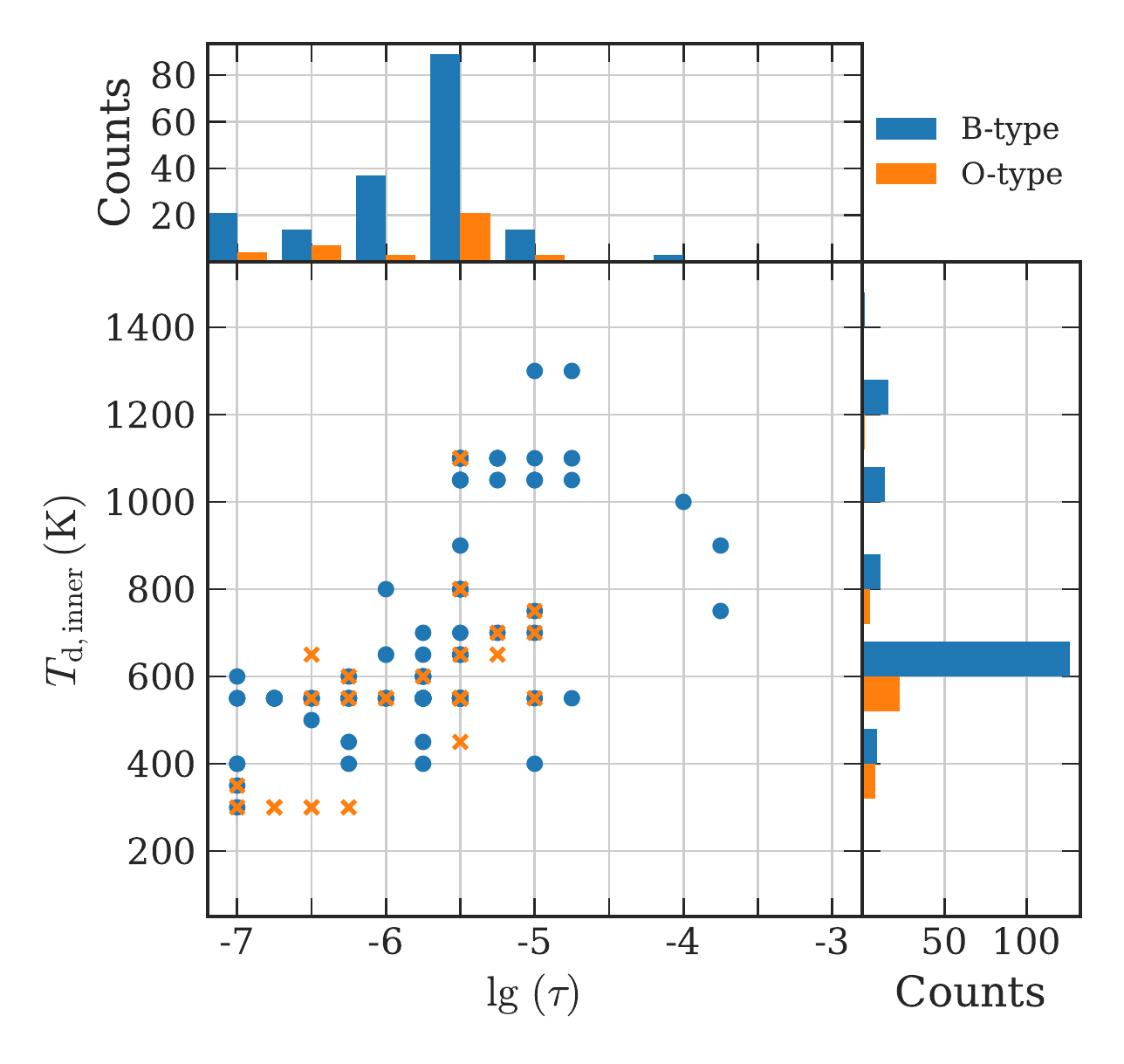}{0.47\textwidth}{}
            }
  \caption{
  Distribution of $T_{\mathrm{d, inner}}$ and $\tau$ for the stars which can be described well by the wind-dust model ($\mathrm{BF} \leq 100$).
  The central panel shows the grid points finally adopted to describe the observational data, and the two histograms at top and right show the distribution of $\tau$ and $T_{\mathrm{d, inner}}$, respectively.
  Blue and orange points (crosses) and bars are used to decode the B-type and O-type stars respectively.
  }
  \label{Fig6:TdTau}
\end{figure}

\begin{figure}
  \gridline{\fig{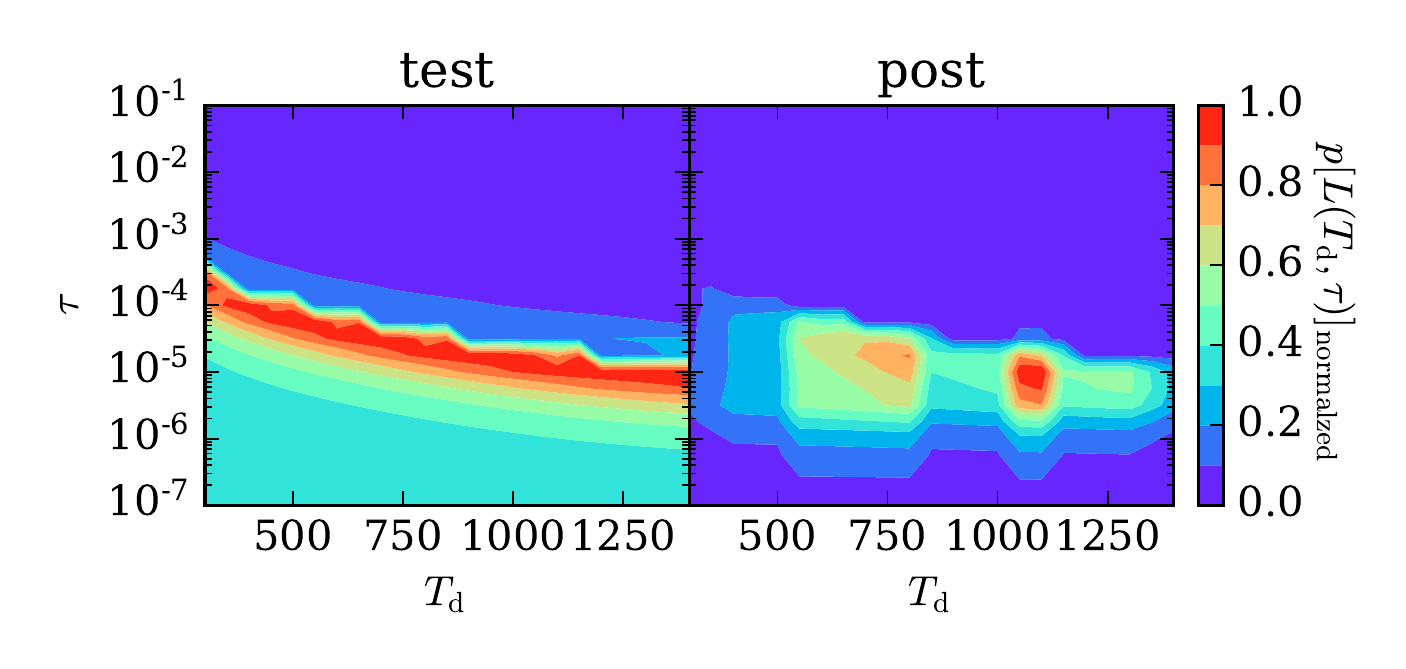}{0.49\textwidth}{}
            }
  \caption{
    The normalized probability of likelihood function ($P[L(T_\mathrm{d}, \tau)] = L(T_\mathrm{d}, \tau)/ \mathrm{max}(L(T_\mathrm{d}, \tau))$) for the second star shown in Figure~\ref{Fig4:SEDmodelExample}. There are multiple maxima in the test run because the lack of long wavelengths observations (left panel), and by applying the prior distribution shown in Figure~\ref{Fig2:Prior}, the post dust parameters are selected.
    }
  \label{Fig7:likelihoodfunction}
\end{figure}

There are 216 stars that could be explained by the wind-dust model, in which 23 of them is fitted better by it.
Figure~\ref{Fig6:TdTau} displays the results of best-fit $T_{\mathrm{d, inner}}$ and $\tau$ for those stars.
The distributions of $T_{\mathrm{d, inner}}$ and $\tau$ from both samples are similar.
As mentioned in Section~\ref{subsec:dustymodel} about the training samples, most of OB star sample are fitted by very low $T_{\mathrm{d, inner}}$ ($\sim 500\,$K) with small $\tau$ ($\sim 10^{-5}$).
Most of the stars in our sample are lack of long wavelengths observations, and there is a very high degree of degeneracy among all the dust parameters, and it is impossible to choose which parameter is better by individual star. 
For example, the second star shown in Figure~\ref{Fig4:SEDmodelExample} only contains data points that reach to $W3$ and $W4$ bands and its IR excess only can be seen in these two bands. 
Before applying the prior probability distribution, there is a very high degree of degeneracy in the `test' runs, i.e., there are multiple maxima in the likelihood function in the parameter space (left panel of Figure~\ref{Fig7:likelihoodfunction}).
The overall likelihood function tends to bias to the parameters on the edge of grids for many of stars under this circumstance. 
With this simple Bayesian method, the parameter degeneracy is eliminated and the optimal parameters can be found in the posterior within the parameter space we set up.
Though the final fitted dust parameters are highly sensitive to the prior probability distribution based on the training sample, these are the best parameters we could estimate for wind-dust model due to the lack of the observational data at long wavelengths.
This incompleteness of wind-dust model grids for some special stars can also be noticed by the 8 stars that couldn't be fitted by either models.
For those targets, a single power-law radiation is not appropriate for explaining their IR excess, and the dust component needed is outside of the current grids. 
More variables, such as another set of dust compositions, are needed, but it is beyond the scope of this study.

The results of fitting as well as stellar parameters are presented in Tables~\ref{tab2:Stars} and \ref{tab3:StarsGOSSS} for the LAMOST and GOSSS sample respectively. 
Because of the asymmetry nature of fitted parameters, the lower and upper bounds of the 68\% (1\,$\sigma$) confidence intervals, $\lambda_{-\sigma}$ and $\lambda_{+\sigma}$, are adopted to show their uncertainties.
Only the stars off the sightlines of dark clouds are presented.
The `Best' column shows which model is the best: 
`S' or `F' is assigned when the stellar wind model with synchrotron radiation or free-free emission is better fitted as $\mathrm{BF} > 100$; `B' is for the case both models work and couldn't be distinguished for which is better as $0.01 \leq \mathrm{BF} \leq 100$; `D' is used when more likely a wind-dust model is needed as $\mathrm{BF} < 0.01$; `U' is used for the stars that none of the model could explain their SED; and `X' is marked when the SED is better fitted by synchrotron radiation in stellar wind but its spectral index is fitted at the lower edge of the grid $\alpha = -10.0$.
There is also an online version of these tables including all of the fitting results.
Meanwhile, Table~\ref{tab4:Samples} shows the samples of various classes.

\begin{splitdeluxetable*}{rrrrcccccBcrrrrrrrrrr}
  \tablecaption{The parameters of the stars with IR excess from LAMOST\label{tab2:Stars}}
  \tablewidth{700pt}
  \tabletypesize{\scriptsize}
  \tablehead{
    \colhead{RA} &        \colhead{DEC} &         \colhead{$T_\mathrm{eff}$} &      \colhead{$\log\,\mathrm{g}$}   & \colhead{$E(B-V)_{\mathrm{D}}$} & \colhead{Luminosity} &  \colhead{err(Luminosity)} & \colhead{Radius} & \colhead{err(Radius)}  & \colhead{Best} &  \colhead{$\mathrm{BF}$} & \colhead{$\alpha_\mathrm{obs}$} & \colhead{$\alpha_{\mathrm{obs}, -\sigma}$} & \colhead{$\alpha_{\mathrm{obs}, +\sigma}$} &   \colhead{$\tau$} &   \colhead{$\tau_{-\sigma}$} &   \colhead{$\tau_{+\sigma}$} &  \colhead{$T_{\mathrm{d}}$} &  \colhead{$T_{\mathrm{d}, -\sigma}$} &  \colhead{$T_{\mathrm{d}, +\sigma}$}   \\
    \colhead{} &        \colhead{} &     \colhead{(K)} &      \colhead{}  &  \colhead{(mag)} &    \colhead{($L_{\sun}$)} &  \colhead{($L_{\sun}$)} & \colhead{($R_{\sun}$)}   &  \colhead{($R_{\sun}$)}   & \colhead{}  &  \colhead{}  &  \colhead{}  &  \colhead{}  & \colhead{}  &  \colhead{} &  \colhead{} &  \colhead{} &  \colhead{(K)} & \colhead{(K)} & \colhead{(K)} 
  }
\startdata
 72.6103 & 40.9471 & 14808 & 3.59 &     0.52 &   1.65e+03 &     7.43e+02 & 2.88e-02 & 2.88e-03 &    B & 1.69e+00 &    -2.1 &          -2.3 &          -1.9 & 1.778e-07 & 1.000e-07 & 1.778e-07 &      550 &      450 &      750 \\
  0.9702 & 57.5972 & 24844 & 4.72 &     0.65 &   3.26e+03 &     1.47e+03 & 1.44e-02 & 1.44e-03 &    S & 1.08e+03 &    -1.0 &          -2.0 &           0.0 & 5.623e-05 & 3.162e-06 & 3.162e-04 &     1100 &      650 &     1300 \\
350.1815 & 49.7993 & 12471 & 4.80 &     0.29 &   5.63e+01 &     2.53e+01 & 7.49e-03 & 7.49e-04 &    S & 1.27e+20 &    -2.9 &          -3.1 &          -2.9 & 5.623e-05 & 5.623e-05 & 1.000e-04 &     1400 &     1350 &     1400 \\
 92.7012 & 13.1835 & 41615 & 4.83 &     0.74 &   1.38e+05 &     6.20e+04 & 3.33e-02 & 3.33e-03 &    B & 2.41e-01 &     1.7 &           1.3 &           1.8 & 3.162e-06 & 5.623e-07 & 5.623e-06 &      550 &      350 &      800 \\
 82.0927 & 39.1771 & 26716 & 3.63 &     0.86 &   3.37e+04 &     1.51e+04 & 3.99e-02 & 3.99e-03 &    B & 1.48e+01 &     1.0 &           0.7 &           1.1 & 3.162e-06 & 5.623e-07 & 5.623e-06 &      550 &      350 &      850 \\
 63.4886 & 50.3225 & 17098 & 3.75 &     1.13 &   2.46e+03 &     1.11e+03 & 2.63e-02 & 2.63e-03 &    F & 6.54e+02 &     1.8 &           0.4 &           2.0 & 3.162e-06 & 1.000e-06 & 1.000e-05 &      550 &      350 &      900 \\
 64.9198 & 50.0809 & 19709 & 3.82 &     1.16 &   2.83e+03 &     1.27e+03 & 2.12e-02 & 2.12e-03 &    B & 4.94e-02 &     2.0 &           1.9 &           2.0 & 5.623e-07 & 1.000e-07 & 5.623e-07 &      400 &      300 &      650 \\
 54.7758 & 50.6611 & 29985 & 6.15 &     1.04 &   4.25e+03 &     1.91e+03 & 1.12e-02 & 1.12e-03 &    B & 2.73e+01 &     1.8 &           1.6 &           1.9 & 3.162e-06 & 5.623e-07 & 5.623e-06 &      550 &      350 &      900 \\
101.0725 &  4.0851 & 13947 & 3.84 &     0.37 &   1.00e+03 &     4.51e+02 & 2.52e-02 & 2.52e-03 &    F & 2.72e+02 &     0.5 &          -0.6 &           0.8 & 3.162e-06 & 1.000e-06 & 1.778e-05 &      550 &      400 &     1000 \\
301.0426 & 30.8007 & 12602 & 4.28 &     0.44 &   3.59e+01 &     1.62e+01 & 5.86e-03 & 5.86e-04 &    S & 6.14e+02 &    -4.8 &          -5.0 &          -3.5 & 5.623e-07 & 1.778e-07 & 1.000e-06 &      600 &      350 &      800 \\
\enddata
\tablecomments{
This table is available in its entirety in machine-readable form.}
\end{splitdeluxetable*}

\begin{splitdeluxetable*}{lrrcccccBcrrrrrrrrrr}
  \tablecaption{The parameters of the stars with IR excess from GOSSS\label{tab3:StarsGOSSS}}
  \tablewidth{700pt}
  \tabletypesize{\scriptsize}
  \tablehead{
    \colhead{Name} &         \colhead{$T_\mathrm{eff}$} &      \colhead{$\log\,\mathrm{g}$}   & \colhead{$E(B-V)_{\mathrm{D}}$} & \colhead{Luminosity} &  \colhead{err(Luminosity)} & \colhead{Radius} & \colhead{err(Radius)}  & \colhead{Best} &  \colhead{$\mathrm{BF}$} & \colhead{$\alpha_\mathrm{obs}$} & \colhead{$\alpha_{\mathrm{obs}, -\sigma}$} & \colhead{$\alpha_{\mathrm{obs}, +\sigma}$} &   \colhead{$\tau$} &   \colhead{$\tau_{-\sigma}$} &   \colhead{$\tau_{+\sigma}$} &  \colhead{$T_{\mathrm{d}}$} &  \colhead{$T_{\mathrm{d}, -\sigma}$} &  \colhead{$T_{\mathrm{d}, +\sigma}$}   \\
    \colhead{} &     \colhead{(K)} &      \colhead{}  &  \colhead{(mag)} &    \colhead{($L_{\sun}$)} &  \colhead{($L_{\sun}$)} & \colhead{($R_{\sun}$)}   &  \colhead{($R_{\sun}$)}   & \colhead{}  &  \colhead{}  &  \colhead{}  &  \colhead{}  & \colhead{}  &  \colhead{} &  \colhead{} &  \colhead{} &  \colhead{(K)} & \colhead{(K)} & \colhead{(K)} 
  }
\startdata
   ALS 12 320 & 29000 & 4.50 &     1.06 &   2.05e+05 &     9.20e+04 & 8.34e-02 & 8.34e-03 &    B &  8.49e+01 &    -6.7 &          -9.9 &          -5.0 & 3.162e-07 & 1.000e-07 & 3.162e-07 &      300 &      300 &      400 \\
   ALS 12 688 & 29000 & 5.00 &     0.92 &   2.16e+05 &     9.72e+04 & 8.58e-02 & 8.58e-03 &    D &  3.29e-05 &     2.0 &          -7.8 &           2.0 & 5.623e-07 & 1.000e-07 & 5.623e-07 &      300 &      300 &      500 \\
      ALS 207 & 25000 & 4.00 &     0.77 &   8.52e+04 &     3.83e+04 & 7.25e-02 & 7.25e-03 &    S &  1.77e+05 &    -3.4 &          -3.8 &          -3.4 & 1.000e-05 & 1.000e-05 & 1.778e-05 &     1400 &     1200 &     1400 \\
     ALS 8272 & 29000 & 4.50 &     0.80 &   8.49e+04 &     3.82e+04 & 5.38e-02 & 5.38e-03 &    S &  5.89e+02 &    -5.6 &          -6.1 &          -5.1 & 3.162e-06 & 5.623e-07 & 3.162e-06 &      400 &      300 &      700 \\
BD +33 1025 A & 50000 & 5.00 &     0.54 &   2.02e+05 &     9.09e+04 & 2.79e-02 & 2.79e-03 &    F &           &     1.2 &           1.1 &           1.2 &           &           &           &          &          &          \\
  BD +39 1328 & 50000 & 5.00 &     0.87 &   5.96e+06 &     2.68e+06 & 1.52e-01 & 1.52e-02 &    F &  2.47e+08 &     2.0 &          -7.4 &           2.0 & 5.623e-07 & 1.000e-07 & 5.623e-07 &      300 &      300 &      500 \\
BD +55 2722 A & 50000 & 5.00 &     0.77 &   5.68e+05 &     2.56e+05 & 4.68e-02 & 4.68e-03 &    D &  1.16e-03 &     0.8 &           0.5 &           0.9 & 1.778e-07 & 1.000e-07 & 1.778e-07 &      300 &      300 &      350 \\
 BD +60 586 A & 28000 & 4.50 &     0.59 &   1.93e+05 &     8.68e+04 & 8.69e-02 & 8.69e-03 &    S &           &    -7.9 &          -8.5 &          -7.5 &           &           &           &          &          &          \\
  BD -08 4623 & 50000 & 5.00 &     1.45 &   7.09e+05 &     3.19e+05 & 5.23e-02 & 5.23e-03 &    S &  1.69e+13 &    -5.8 &          -5.9 &          -5.8 & 1.000e-05 & 5.623e-06 & 1.000e-05 &      300 &      300 &      350 \\
 CPD -26 2716 & 50000 & 5.00 &     0.71 &   1.40e+07 &     6.30e+06 & 2.32e-01 & 2.32e-02 &    F &  3.05e+16 &     2.0 &           1.9 &           2.0 & 5.623e-07 & 1.000e-07 & 5.623e-07 &      300 &      300 &      550 \\
\enddata
\tablecomments{
This table is available in its entirety in machine-readable form. 
}
\end{splitdeluxetable*}

\begin{deluxetable*}{lccclccclccccc}
\tablecaption{The samples of various classes\label{tab4:Samples}}
\tablewidth{0pt}
\tabletypesize{\scriptsize}
\tablehead{    \colhead{}   & \multicolumn{3}{c}{Initial Sample}  & \colhead{} & \multicolumn{3}{c}{IR excess} & \colhead{} & \multicolumn{5}{c}{Best Model}        \\
\cline{2-4}
\cline{6-8}
\cline{10-14}
\colhead{} & \colhead{in DC}        &        \colhead{off DC}         & \colhead{all} & \colhead{} & \colhead{in DC} & \colhead{off DC} & \colhead{all} & \colhead{} & \colhead{Wind: Free-Free}             & \colhead{Wind: Synchrotron} & \colhead{Both} & \colhead{Wind-dust} & \colhead{Unknown}
}
\startdata
B-type & 1669 & 5634 & 7303 & &  407  &  532  & 939  & & 178 & 169 & 159 & 19 & 7\\
O-type & 214  & 301  & 515  & &  90   &  118  & 208  & & 39  & 40  & 34  & 4  & 1\\
All &    1883 & 5935 & 7818 & &  497  &  650  & 1147 & & 217 & 209 & 193 & 23 & 8\\
\enddata
\tablecomments{`in DC': inside the sightline of Dark Clouds. `off DC': off the sightline of Dark Clouds. `Both': both stellar wind and wind-dust models could explain the observations. `Unknown': none of the two could explain the observations.
}
\end{deluxetable*}


\section{Discussion}
\label{sec:Discussion}

\subsection{Be-stars}

Be-stars are non-supergiant B-type stars with at least one Balmer line emission.
For classical Be-stars, which are fast-rotating main-sequence B-type stars, the outflowing material forms a gaseous dust-free Keplerian circumstellar disk \citep{rivinius_classical_2013}. 
The IR excess is often observed on them and it is mostly due to free-free emission from that disk.
For this Be phenomenon, its emission is most likely to be a power-law with a spectral index $\alpha \in [0.6, 2]$ \citep{Klement_BeCSM_2017}.
\citet{Carciofi_BeDisks_2006} predicted that for fully ionized gaseous disks near Be-stars, the IR excess should show up at wavelengths $\leq 1\,\micron$ for disks inclinations $\leq 60$\arcdeg \, and at $\sim 10\,\micron$ for edge-on condition.

\citet{rivinius_classical_2013} summarizes the $H_{\alpha}$ profile under the present classical Be-star model with a gaseous disk. 
The double peaks around the core correspond to the edge-on case.
The single peak represents the pole-on condition, in which the wing-absorption might exist due to the strong stellar photospheric absorption.
All these line profiles are observed in our LAMOST spectra.
Similar to samples from \citet{siebenmorgen_far-infrared_2018}, there are many stars in our sample that contain far-IR emission which only appears at wavelengths $> 10\,\micron$.
Most of them also have double-peak $H_{\alpha}$ emission line, which well support the edge-on condition of the model for classical Be-star with gas disk.

Among the 532 stars with IR excess from the LAMOST-OB catalog off dark clouds sightlines and which are identified as B-type stars, 153 sources exhibit a clear $H_{\alpha}$ emission line ($\sim 29\%$) indicating that they are Be-stars.
For 109 sources of them, the SED could be best fitted by a stellar wind model ($\mathrm{BF} > 100$) and free-free emission works for 57 of them.
It leads to a relatively higher proportion ($109/153 \sim 71\%$) of successful wind model than for the entire sample of B-type stars ($\frac{178+169}{532} \sim 65\%$), which is consistent with the scenario that a Be-star usually has an ionized gaseous disk.

\subsection{Possibility of Synchrotron Radiation}

For the stars whose SED can be fitted by a power-law but with $\alpha < -0.1$, the infrared excess can hardly be attributed to free-free emission which in general decreases with wavelength. 
On the other hand, such an SED that increases with wavelength resembles the synchrotron radiation. 
Although some early-type in particular B-type stars are found to have strong magnetic field, there must be some mechanism to obtain relativistic electrons if synchrotron radiation is to occur. 
\citet{Synchrotron_White1985ApJ...289..698W} proved that electrons can be accelerated to relativistic energies by chaotic stellar winds in hot stars, and \citet{Shchekinov&Sobolev2004_SynchrotronMassivestars_A&A...418.1045S} argued that the interaction of stellar wind with the surface of a circumstellar disk can result in the acceleration of relativistic radiation.

\begin{figure}
  \gridline{\fig{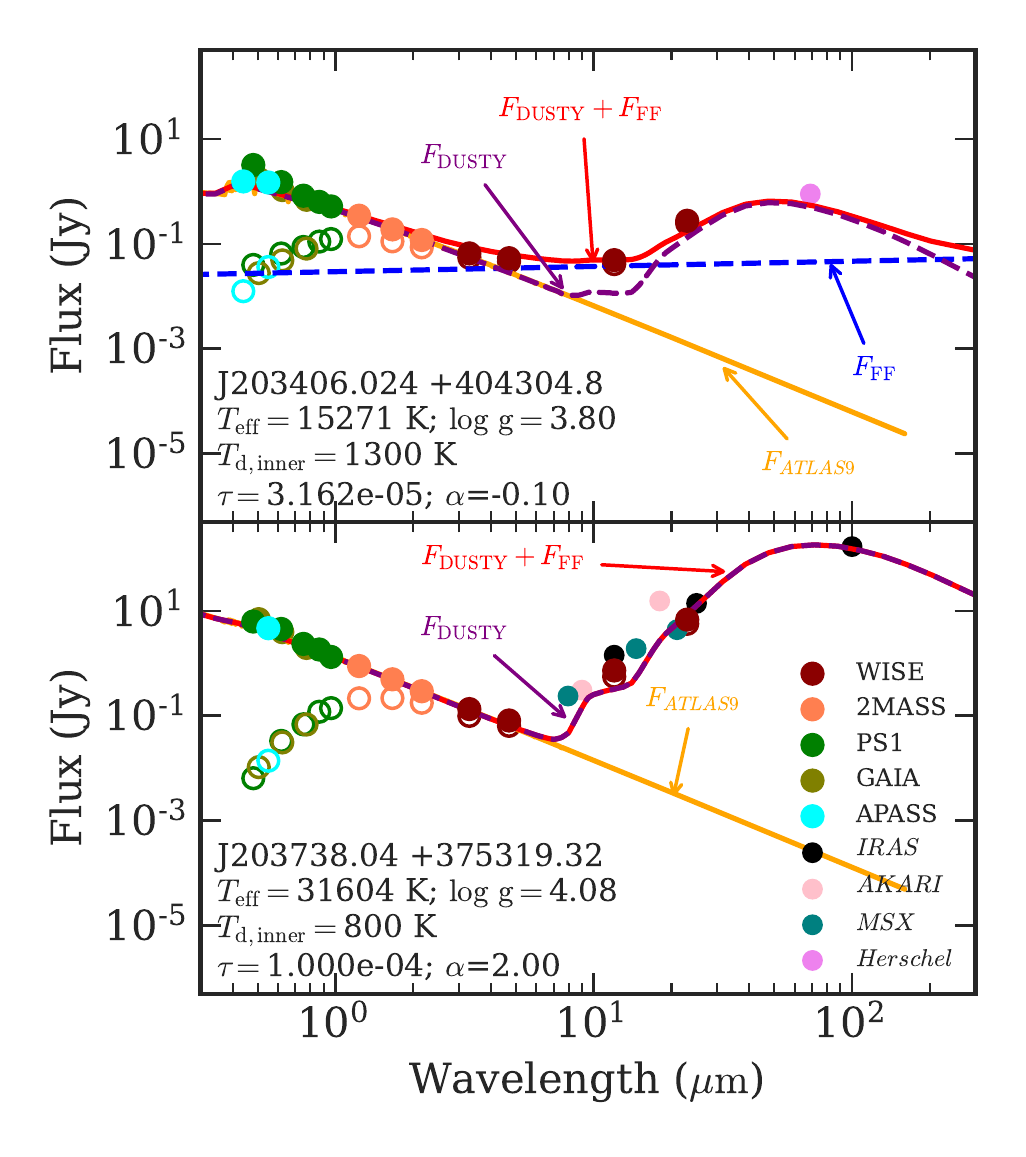}{0.48\textwidth}{}
            }
  \caption{
  The wind-dust model fitting of the two stars with long wavelengths observations from \emph{IRAS}, \emph{AKARI} and \emph{Herschel} measurements up to 100\,\micron, where the turn-over points appear in both of them. Observational data points are marked in color, and all lines follow the convention in Figure~\ref{Fig4:SEDmodelExample}. In the lower panel, the stellar wind component would not help in the fitting and thus is not included in the best-fit model.}
  \label{Fig8:SEDExampleSpecial}
\end{figure}

27 objects (the training sample in dust model discussed in Section~\ref{subsec:dustymodel}) are detected at relatively longer wavelengths by either \textit{IRAS}, \textit{AKARI} or \textit{Herschel} with observations at $\lambda \ge 60\,\micron$, and 10 of them are out of the dark clouds sightlines. For all these 27 stars, their infrared excess can be fitted by a steep power law ($\alpha < -0.1$) except one with $\alpha \sim 0.2$, and several with a plausible turn-over point at $\sim 100\,\micron$ (see Figure~\ref{Fig8:SEDExampleSpecial}).
Besides, the search for a counterpart in the NVSS catalog \citep{NVSS_catalog_Condon1998AJ....115.1693C} within 5\,\arcsec\  for each star results in only 3 stars in spite that the extrapolation of the power law predicts an intensity at 21\,cm greatly higher than the sensitivity of the NVSS survey. 
In addition, for those 3 stars, the NVSS observed flux is too low comparing with the predictions from synchrotron radiation starting from IR wavelengths (more than 3 orders of magnitude lower in the unit of Jy), so that their radio emission is more likely to be originated from circumstellar dust emission or other mechanisms.
In the total sample of 650 stars, the IR excess of 275 (24\%) sources can be fitted by the synchrotron radiation and 209 (18\%) of then even favor this model.
But it is difficult to explain such large proportion deserving synchrotron radiation.
Since the dust in dark cloud is usually cold and emits in far-infrared, the infrared excess originates very possibly from the dark cloud, and the variation of the SED within the studied wavelength range is caused by the difference in temperature.
Alternatively, the wind-dust model is a better explanation for many of them.
Thus, the final solution lies in more brightness measurements at longer wavelengths which can help distinguish these models.

\subsection{Possibility of Debris Disk}

Disk is one of the ubiquitous dynamic structures in the universe. 
As pre-main-sequence stars, Herbig Be stars with protoplanetary disks have now been studied extensively. 
Dusty debris disks might also exist around main-sequence massive stars as they evolved. 
\citet{DebrisDisks_Roberge2008ApJ...676..509R} presented SED for 16 nearby main-sequence massive stars including one Be-star (HD142926), one early F-type and 14 A-type stars. Both of the \textit{Spitzer}/MIPS/24 and 70\,\micron\ photometry are included in their observations. 
It is found that the mid-IR excess in that Be star does not like the classical Be stars, whose fluxes should be a power-law decline with increasing wavelength. 
A debris disk model with blackbody dust grains or 1\,\micron\ silicate grains can both fit the observational data but with different parameters such as dust temperature, indicating a possible existence of the dusty debris disk around it. 
For a cold dust disk model, its SED also shows a continuous increase in fluxes from 10 to 100\,\micron , which is very similar to our current IR observation.

It is impossible to distinguish the disk structure from the spherical circumstellar dust described in this work (Section \ref{subsec:dustymodel}) until a directly resolved image is acquired.
\citet{siebenmorgen_far-infrared_2018} performed near-IR high-contrast imaging of three O-type stars with far-IR excess in their sample, and they didn't find any significant disk structures except the stellar halo, which might be the scatter-light from disks.  
Also, it is basically impossible to reach a reasonable constraint on the disk model due to the lack of observation data at longer wavelength, and there is no direct evidence that debris disks exist around main-sequence OB stars. 
Therefore, further research on the disk model is not conducted in this work.

\subsection{Dust Source}
\label{subsec:DustSource}

Normally, the strong stellar wind of the star is accompanied by violent material ejection, which then brings considerable circumstellar matter.
However, the harsh environment near OB stars brought challenges to the survival of circumstellar dust.
The strong stellar wind blows away dust, and the Poynting-Robertson drag also causes dust near inner radius to lose angular momentum and fall into the stellar atmosphere \citep{draine_physics_2011}.

From the grids generated by DUSTY (Section~\ref{subsec:dustymodel}), a parsec-scale dusty sphere with a very small optical depth ($\tau \leq 10^{-5}$) and rather low dust temperature ($T_{\mathrm{d, inner}} \sim 500\,$K) can best explain the observational data.
A possible condition should be: as an OB star born in its molecular cloud, the strong stellar wind quickly blows away the surrounding cloud and a huge structure was constructed, making a parsec-scale dusty envelope.
That is to say, this envelope is the molecular cloud blown larger by the stellar wind.
Thus, it has the same chemical composition and dust density distribution as this cloud, which are both represented in our wind-dust model (see Section~\ref{subsec:dustymodel}).
Far from the central star, the temperature is low, and the optical depth is small, which looks like a dusty halo as suggested by \citet{siebenmorgen_far-infrared_2018}.
This dusty circumstellar halo, together with the photosphere and stellar wind radiation inside, can interpret the observational SED.

This circumstellar halo strucutre is similar to the scenario of a Young Stellar Objects (YSO) growing in a dark molecular cloud. 
\citet{YSO_Molinari2008A&A...481..345M} described how the SEDs of massive YSOs evolve in the star forming region. 
They adopted both DUSTY code and 3-dimentional model by \citet{RT_Whitney2003ApJ...591.1049W} to estimate the SED and fit the observational data points focusing on wavelengths from $\sim 10\,\micron$ to $> 1000\,\micron$, c.f., Figures in Appendix A in \citet{YSO_Molinari2008A&A...481..345M}.
Although their objects are massive YSO, their observational SEDs are very similar to our samples of main-sequence OB-type stars requiring dust components: there is a sharp increase of IR excess to $100\,\micron$. 
Their fitted DUSTY parameters are also similar to our results in the wind-dust model. 
Besides, shown in Figure~\ref{Fig8:SEDExampleSpecial}, there are turn-over points at $100\,\micron$ for these two stars staying in dark clouds sightlines, which makes it even closer to the YSO in dark cloud scenario.
Therefore, the dust components in the IR excess actually reflects the conditions of the birthplaces of these massive stars: as they grow extremely fast, the dark clouds as the birthplaces still surround them when they are  already in the main-sequence phase.
For those stay out of the currently identified dark clouds sightlines, they might stay in a very small or faint dark cloud that has not been identified.

\subsection{Dependence of infrared excess on stellar parameters}

\begin{figure}
  \gridline{\fig{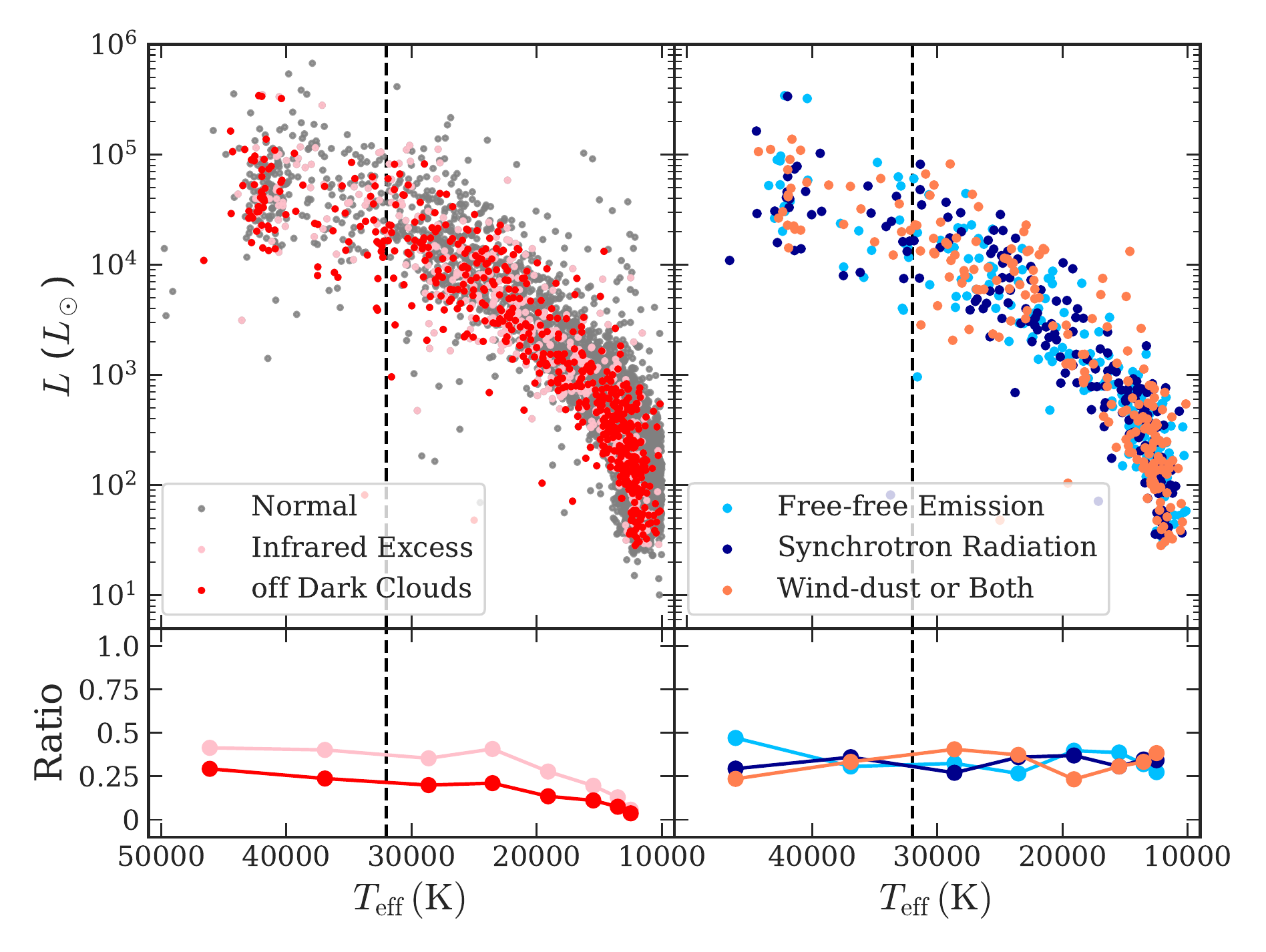}{0.49\textwidth}{}
            }
  \caption{
    Dependence of the IR excess and its model on stellar effective temperature and luminosity.
    The distribution of the stars with (pink dots) and without IR excess (grey dots) in the HR diagram are shown in the upper-left panel, and the stars with IR excess off the dark clouds are shown in the red dots.
    The change of the proportion with infrared excess is displayed in the lower-left panel.
    The right column is for the SED that is best fitted by free-free emission (light blue dots), synchrotron radiation (dark blue dots) of stellar wind model, or could be fitted by wind-dust model (orange dots).
    The $T_\mathrm{eff} = 32,000\,\mathrm{K}$ is shown in black lines to divide the O-type and B-type stars in our sample. 
  }
  \label{Fig9:HRD}
\end{figure}

The LAMOST sample is taken to investigate the dependence of IR excess on stellar parameters, for which the stellar parameters are available.
The GOSSS sample provides no stellar parameters and is dropped off.
The stellar luminosity is derived with the effective temperature from the LAMOST spectroscopy and the radius calculated by Equation \ref{Eq1:Fobs}.
Given that the errors of the effective temperature and the radius are both 10\%, the luminosity is with $\sim 45\%$ uncertainty.
Figure~\ref{Fig9:HRD} shows the distribution of stars with IR excess and best models in the Hertzsprung-Russell diagram (HR diagram).
The percentage with IR excess increases apparently with stellar effective temperature ($T_{\mathrm{eff}}$), and the luminosity since the objects are all dwarfs.
For B-type stars , when $T_{\mathrm{eff}} > 20,000\,$K, more than 25\% of them presents IR excess, while this percentage reduces to $<5\%$ at $T_{\mathrm{eff}} \sim 10,000\,$K.
This trend agrees with the origin of the IR excess that is mainly stellar wind since luminous OB stars generally have strong stellar wind.
For the mechanisms explaining the IR excess (right panels of Figure~\ref{Fig9:HRD}), all three cases generally stay at the same level at different $T_{\mathrm{eff}}$, except for slightly changing trends in lower temperature end.
Free-free emission from ionized stellar wind and synchrotron radiation are more common for stars with higher $T_{\mathrm{eff}}$, rising to $\sim 40\%$ at $T_{\mathrm{eff}} \sim 20,000\,$K.
Meanwhile, the dusty structures are more likely to survive around stars with lower temperatures, thus the proportion of sources requiring dust component reaches to $\sim 40\%$ on the low temperature end with a small increase.
However, since the variations in both trends are just above the Poisson uncertainties ($\frac{\sqrt{N_\mathrm{bin}}}{N_\mathrm{total}} \sim 5\%$ in ratio) in each bin, those two trends are questionable.
Thus, we conclude that there is no obvious systematic change of the mechanism for infrared emission with stellar parameters.

\section{Summary}
\label{sec:Summary}

The infrared excess of OB stars is systematically studied based on the largest OB star catalog with stellar parameters from the LAMOST survey and the GOSSS O-type star sample.
After a precise extinction-correction with the intrinsic color indexes from our previous work \citep{deng_intrinsic_2020}, IR excess are identified by comparing their spectral index in the infrared SED with photospheric model in a forward modeling approach.
It is found that 939 and 208 stars show infrared excess among the 7303 B-type and 515 O-type stars, respectively.
To better analyze the circumstellar condition of them, 407 B-type and 90 O-type stars in the dark clouds sightlines are eliminated, which leaves 532 ($\sim 7\%$) and 118 ($\sim 23\%$) stars respectively with true circumstellar infrared excess.
Afterwards, the observational SED from optical bands (\textit{Gaia}, PS1 and APASS) to infrared (2MASS, \textit{MSX}, \textit{Spitzer} and \textit{WISE}) is interpreted by synchrotron radiation or free-free emission in stellar wind or together with dust thermal radiation.
Bayes Factors ($\mathrm{BF}$) are computed for both models to quantitatively compare which one fitting observations better.
The IR excess in one third of the OB-type stars ($\sim 33\%$) can be better explained by free-free emission in ionized stellar wind, 
another one third ($\sim 32\%$) of them is explained better by synchrotron radiation, both models work well for other 193 stars ($\sim 30\%$), while wind-dust model works better for only 23 sources ($\sim 4\%$) and 8 stars ($\sim 1\%$) couldn't find a proper model.
For those objects that wind-dust model could fit the observations ($\sim 34\%$ of the total sample), a parsec-scale dusty envelope with a low dust temperature and exceedingly small optical depth is identified, which implies a large-scale circumstellar dust halo possibly originated from the birthplace cloud.

\acknowledgments We are grateful to Prof. Aigen Li from University of Missouri, Prof. Chao Liu from National Astronomical Observatories, Chinese Academy of Sciences, Yanjun Guo from Yunnan Observatory, Chinese Academy of Science, and Profs. Jian Gao and Hai-Bo Yuan from Beijing Normal University for very helpful discussions. 
We also thank the anonymous referee for the suggestions.
This work is supported by the NSFC projects 12133002 and 11533002, National Key R\&D Program of China No. 2019YFA0405503 and CMS-CSST-2021-A09.
 This work has made use of data from LAMOST, PS1, APASS, \textit{Gaia}, 2MASS, \textit{MSX}, \textit{Spitzer}, \textit{WISE}, \textit{IRAS} and \textit{Herschel}.
 
\bibliography{IEOB}{}
\bibliographystyle{aasjournal}


\end{CJK*}
\end{document}